\documentclass[11pt]{article}

\usepackage[usenames,dvipsnames,table]{xcolor}
\usepackage[utf8]{inputenc}

\pdfoutput=1 
\usepackage{jheppub} 

\usepackage{graphicx}
\usepackage{amsmath, amssymb}
\usepackage{comment}

\usepackage{mathrsfs}

\usepackage{framed}
\definecolor{shadecolor}{rgb}{0.90,0.90,0.90}

\usepackage{mathtools}

\def\be{\begin{eqnarray}}
	\def\ee{\end{eqnarray}}
\def\bea{\begin{eqnarray}}
	\def\eea{\end{eqnarray}}

\newcommand{\Tr}{\mbox{Tr}}

\newcommand{\beal}{\begin{equation}
		\begin{aligned}}
		\newcommand{\eeal}{\end{aligned}
\end{equation}}
\newcommand{\bem}{\begin{multline}}
	\newcommand{\eem}{\end{multline}}

\def\U{\text{U}}
\def\SU{\text{SU}}
\def\USp{\text{USp}}

\def\E{\text{E}}

\def\cC{\mathfrak{C}}
\def\CC{\tilde{C}}
\def\CCC{\tilde{\mathcal{C}}}
\def\ccA{C}
\def\ppsi{\mathbf{\Psi}}

\makeatletter
\gdef\@fpheader{\break}
\makeatother

\begin{document}
    \begin{flushright}
    YITP-SB-2026-06
    \end{flushright}
    
	\title{On non-relativistic integrable models and $4d$ SCFTs
    }
	
	\medskip
	
	\author[1]{Rotem Ben Zeev,\!}
	\author[2]{Anirudh Deb,}  
    \author[3,4]{Hee-Cheol Kim,}
	\author[1]{and Shlomo~S.~Razamat}

	\medskip
	
	\affiliation[1]{Department of Physics, Technion, Haifa, 32000, Israel}

    \emailAdd{rotem.b@campus.technion.ac.il} \emailAdd{anirudh.deb@stonybrook.edu}
    \emailAdd{heecheol@postech.ac.kr}
    \emailAdd{razamat@technion.ac.il}

	\affiliation[2]{C.N.Yang Institute for Theoretical Physics, Stony Brook University, Stony Brook,NY 11794-3840,USA}

\affiliation[3]{Department of Physics, POSTECH, Pohang 37673, Korea}
\affiliation[4]{Asia Pacific Center for Theoretical Physics, POSTECH, Pohang 37673, Korea}


	\abstract{We elaborate on the relation between the generalized Schur index of ${\cal N}=2$ SCFTs in four dimensions and the non-relativistic limit of the elliptic Ruijsenaars-Schneider model. In particular we discuss explicitly how to express generalized Schur indices of theories of class ${\cal S}$  in terms of elliptic Jack functions. For example, in the $A_1$ case the indices are given naturally in terms of eigenfunctions of the Lam\'{e} equation. We use the  expression in terms of eigenfunctions to further check the recent observation that the generalized Schur indices of different theories in the Deligne-Cvitanovi\'{c} series can be mapped onto each other. This mapping implies non trivial identities on unrefined sums of eigenfunctions of non-relativistic elliptic Calogero-Moser models associated to different root systems. We claim then that the non-relativistic limits of various integrable models give rise naturally to generalized Schur-like limits of classes of ${\cal N}=1$ SCFTs. As an example we discuss
  the relation of the Inozemtsev model, the non relativistic limit of the van Diejen model, and compactifications of the rank $Q$ E-string theory. We argue that in general the ``Schur index'' of ${\cal N}=1$ $4d$ SCFTs can be understood as being related to the free fermionic limit of a non-relativistic integrable model. }

	\maketitle

    \section{Introduction}

    Supersymmetric partition functions have proven to provide us with a very useful tool to probe interesting physics by penetrating through strong coupling barriers. Being independent of certain continuous deformations they divide the space of supersymmetric theories into universality classes connected by such deformations.

\begin{figure}
 \center
 \includegraphics[width=0.8\textwidth]{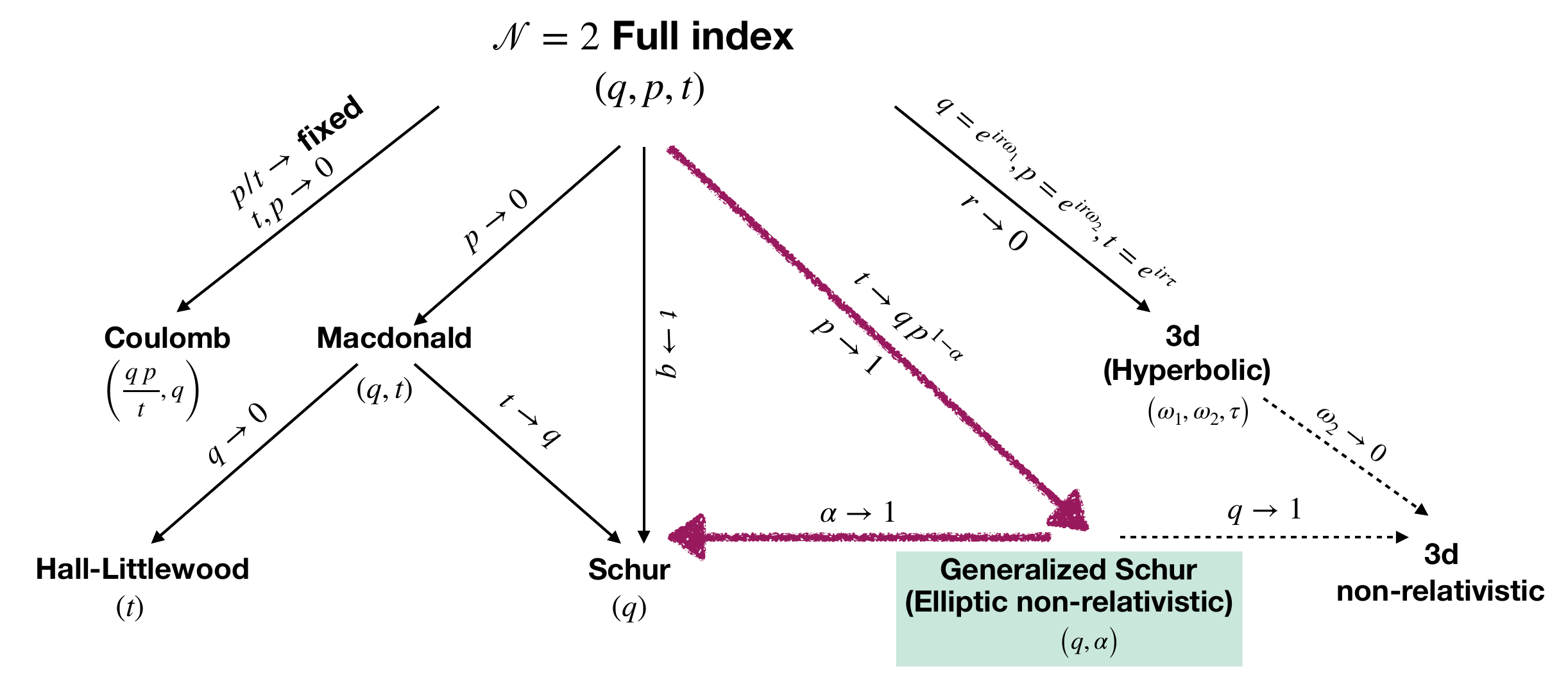} 
 \caption{(A partial) map of different general degeneration limits of the ${\cal N}=2$ superconformal index. On the left side of we have various supersymmetry enhancing limits (Coulomb/Hall-Littlewood/Macdonald/Schur). The main focus of the paper will be the generalized Schur partition function, 
 {\it aka} elliptic non-relativistic limit of the RS model. Missing from this map are some further non-relativistic limits.}
 \label{F:MapOfLimitsN2}
 \end{figure}

     Let us consider $4d$ ${\cal N}=2$ SCFTs. The different supersymmetric indices for the ${\cal N}=2$ SCFTs form a natural hierarchy \cite{Kinney:2005ej,Gadde:2011uv,Rastelli:2014jja}. At the top of the hierarchy is what is usually called the {\it full} index depending generally on three parameters, $q$, $p$, and $t$, as well as on parameters corresponding to the global symmetry of the given theory,
     \be\label{eq:FullIndex}
     {\cal I}_{p,q,t}=\Tr_{{\mathbb S}^3}(-1)^F p^{\frac12(\Delta+2j_1-2R-r)}q^{\frac12(\Delta-2j_1-2R-r)}t^{R+r}\, e^{-\beta'(\Delta-2j_2-2R+r)}\,,
     \ee where the trace is taken over the Hilbert space in radial quantization, $\Delta$ is the scaling dimension, $R$ is the Cartan generator of $\SU(2)_R$ R-symmetry, $r$ is the generator of $U(1)_r$ R-symmetry, and $j_i$ are the Cartan generators of the $\SU(2)_1\times \SU(2)_2$ isometry of ${\mathbb S}^3$.
     All the other indices can be obtained by some degeneration limits of the full index. See Figure \ref{F:MapOfLimitsN2} for a partial map of the hierarchy of indices.

	All of these degeneration limits are well defined for {\it any} ${\cal N}=2$ SCFT. Some of the limits depicted in Figure \ref{F:MapOfLimitsN2} were studied in detail in the literature starting with \cite{Gadde:2011uv}. Let us briefly discuss some of these. The Coulomb limit is solely determined by the Coulomb  branch operators. As these are freely generated for theories of class ${\cal S}$ \cite{Gaiotto:2009we,Gaiotto:2009hg} they contain very little information.\footnote{One can consider various discrete gaugings relaxing the constraint of free generation of the Coulomb branch \cite{Argyres:2017tmj,Bourton:2018jwb,Argyres:2018wxu,Argyres:2024uuc}.}
     The remaining three limits on the left side of the map are still simple on one hand but contain a lot of non trivial information about the SCFT on the other hand. In particular,
     these limits are very easy to compute for the vast landscape of theories of class ${\cal S}$, ${\cal N}=2$ SCFTs which can be obtained as punctured Riemann surface compactifications of the six dimensional $(2,0)$ theories. The general structure of the index for such theories is \cite{Gadde:2009kb,Gadde:2011uv,Gaiotto:2012xa},
\begin{equation}\label{eq:generalformoftheindex}
		\mathcal{I}_{\mathfrak{g},\mathfrak{s}}(\left\{\mathbf{a}_I\right\})=\sum_{\lambda} B_\lambda \left(C_{\lambda}\right)^{2\mathfrak{g}+\mathfrak{s}-2} \prod_{I=1}^\mathfrak{s}\psi_{\lambda}(\mathbf{a}_I)~.
\end{equation} Here ${\mathfrak g}$ and ${\mathfrak s}$ are the genus and the number of (maximal) punctures respectively, and ${\bf a}_I$ are fugacities associated to the symmetries of the maximal punctures. The factor $B_\lambda$ encodes contributions of non-maximal punctures. The functions $\psi_\lambda(\mathbf{a})$ are equal to Macdonald/Hall-Littlewood/Schur polynomial (up to $\lambda$-independent functions) in the corresponding limits. These polynomials are very well studied and thus these limits of the index are easily computable. Physically these three limits count certain (at least) quarter BPS operators. The index can be defined as an ${\mathbb S}^3\times {\mathbb S}^1$ partition function with twisted boundary conditions, and thus one can consider taking the radius of the circle to zero obtaining ${\mathbb  S}^3$ partition function of the circle reduced theory. This is the $3d$ limit appearing on the right side of the map of Figure \ref{F:MapOfLimitsN2}.

Note that the full index can be defined as a weighted counting of certain supersymmetric states/operators in the theory. Coulomb/Macdonald/Hall-Littlewood/Schur limits still have a counting interpretation. In particular, these indices can be written as a Taylor series in some of the parameters coefficients of which are integers. On the other hand the $3d$ limit loses a natural counting interpretation and has no natural expansion with integer coefficients.\footnote{Note however, that even the $3d$ partition function can be directly related to $3d$ counting problems by constructing the ${\mathbb S}^3$ geometry in terms of $\text{Disc}\times {\mathbb S}^1$ building blocks \cite{Pasquetti:2011fj,Beem:2012mb}. Moreover, expressions of the index of the form \eqref{eq:generalformoftheindex} can be related \cite{Nishioka:2011dq} to the star-shaped structure of dimensionally reduced  theories \cite{Benini:2010uu}.}

The {\it full} index has similar expansion to \eqref{eq:generalformoftheindex} with $\psi_\lambda(\mathbf{a})$ equal to (up to $\lambda$-independent function) eigenfunctions of the Ruijsenaars-Schneider (RS) model \cite{Gaiotto:2012xa}.\footnote{
This is an example of the many deep interconnected relations of integrable models, CFTs (mainly with eight supercharges), and related constructions; see {\it e.g} \cite{Gorsky:1993pe,Donagi:1995cf,Gorsky:1995zq,Nekrasov:2009uh,Nekrasov:2009rc,Alday:2009aq,Gadde:2009kb,LeFloch:2020uop}.
} The type of the  RS model is determined by the type of the $6d$ $(2,0)$ model compactifications of which we are studying. The RS model is an example of a {\it relativistic} integrable model. The relativistic nature refers to the fact that the Hamiltonians of these models are given in terms of finite difference operators rather than differential ones.\footnote{Note that the relativistic/non-relativistic here refers to the integrable model and not to the underlying $4d$ QFTs which are always relativistic. We refer the reader to \cite{Ruijsenaars1999} for a nice overview of the integrable models nomenclature. For a discussion of supersymmetric indices of non-relativistic QFTs see \cite{Nakayama:2008qm}.} For example, in the one parameter ($A_1$) case, the basic Hamiltonian when acting on some test function $f(z)$ is a linear combination of shifted evaluations of the function, $f(p^{\pm\frac12}z)$. It is natural then to consider the limit of $p\to 1$ whence such finite shifts will become derivatives. Such limits are called {\it non-relativistic} limits of the integrable models. For these limits to be well defined typically other parameters also need to be scaled in some way. For example, in the case of the elliptic RS model one has also to scale $\frac{q p}t\to p^\alpha$ with fixed $\alpha$ while taking $p\to 1$ in order to obtain a well defined non-relativistic limit. 
We will discuss this in detail in the next sections. The special case of $\alpha=1$ is the Schur index \cite{Gadde:2011ik,Gadde:2011uv}.
In the context of limits of the ${\cal N}=2$ supersymmetric index such a non-relativistic limit was considered  in \cite{Cecotti:2015lab,Deb:2025ypl}. In \cite{Deb:2025ypl} the limit  was referred to as the generalized Schur partition function. The goal of this paper is first to elaborate on the connection between non-relativistic limit of the RS model and the physics of the generalized Schur partition functions of ${\cal N}=2$ SCFTs, as well as to extend such connections to other integrable models and ${\cal N}=1$ SCFTs.

\begin{figure}
 \center
 \includegraphics[width=0.8\textwidth]{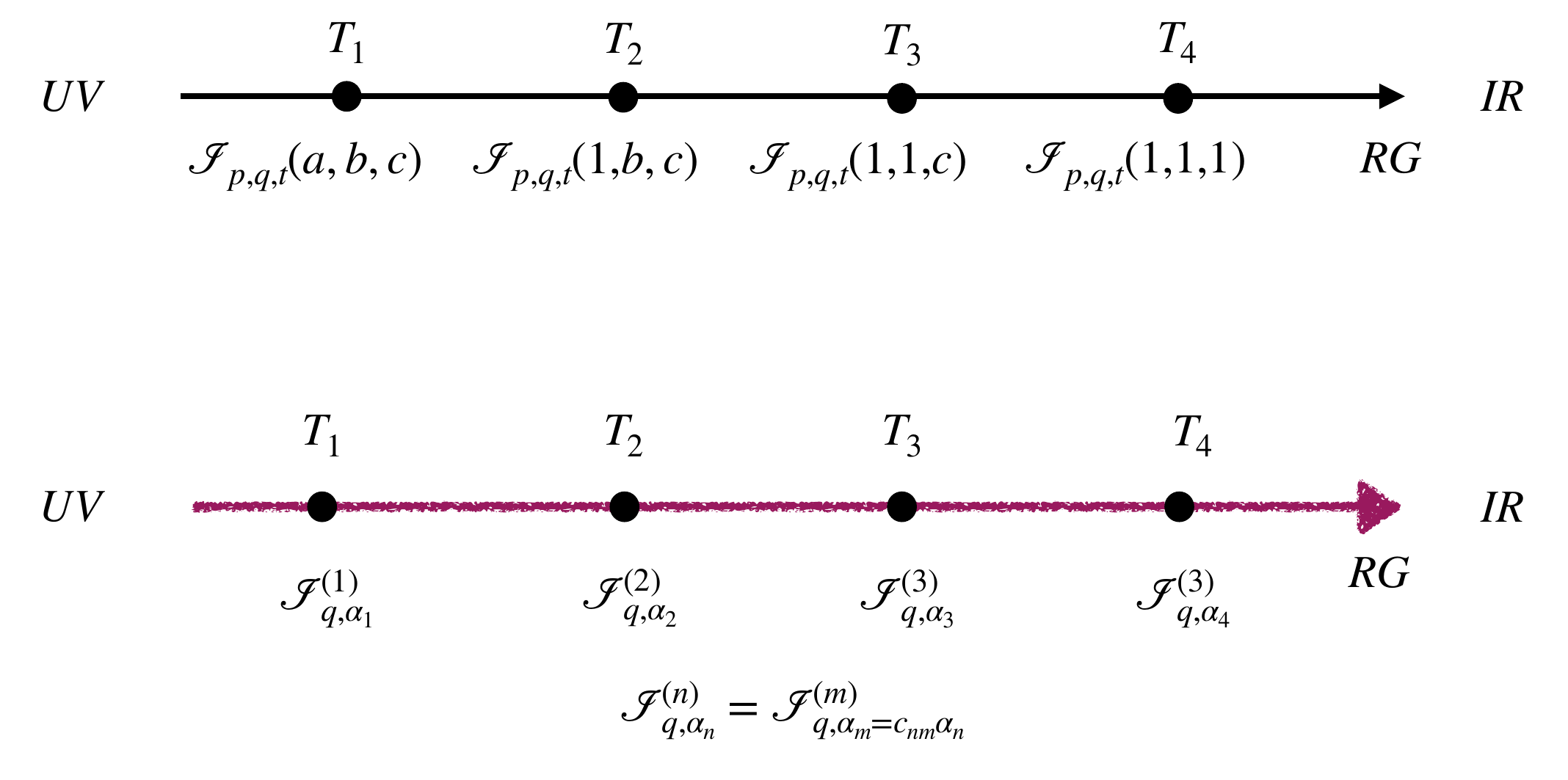} 
 \caption{On the top is the relation of indices of theories connected by flows not breaking $\U(1)_r$ symmetry. The indices for theories down the flow are obtained from indices of theories up the flow by specializing parameters. In particular the index down the flow is also a specialized index of a theory up the flow. On the bottom 
 a different relation between indices which can be observed in some flows breaking $\U(1)_r$ (mass terms and/or Coulomb branch flows). The generalized Schur partition function of different theories are related by a non-trivial map of parameters $\alpha$: the couplings of the associated non-relativistic integrable model}
 \label{F:GeneralizedSchurFlows}
 \end{figure}

The generalized Schur partition function has certain surprising properties. To explain these, let us first review partition functions of theories related by various deformations. Consider general ${\cal N}=1$ SCFT with $\U(1)_r$ R-symmetry. First, assume that the deformation preserves $\U(1)_r$ R-symmetry used to compute a non-trivial index of the UV theory. These deformations can be superpotential deformations or turning on vacuum expectation values (vevs).
In the case of superpotential deformation the index of the IR theory is given  as a specialization of the UV index switching off  the fugacities for symmetries broken by the deformation.
In the case of a vev, switching off the fugacities for symmetries broken by the vev leads to a pole divergence of the index and the procedure to obtain the index of the IR theory is adjusted to compute the residue of this pole and divide away the contributions of the Goldstone fields \cite{Gaiotto:2012xa}. Importantly, the index of the IR theory is also the index of the UV theory albeit with not the most generic parameters and not utilizing the IR superconformal R-symmetry.

Next, consider deformations which break the $\U(1)_r$ symmetry which renders the UV index finite. Again, this can be either a superpotential or a vev deformation.  Then the indices of the UV and the IR theories are not continuously related to each other as the computation of the index requires an existence of an R-symmetry \cite{Festuccia:2011ws}. The IR CFT might have an {\it emergent} $\U(1)_r$ symmetry using which one can compute the index of the IR SCFT, but it is not directly related in general to computations done in the UV. 

In the ${\cal N}=2$ setup we have $\SU(2)_R\times \U(1)_r$ R-symmetry. We can consider starting from an SCFT and compute the full index. Then we can turn on superpotential terms linear in moment map operators and/or turn on vevs for Coulomb branch operators. These deformations preserve $\SU(2)_R$ symmetry but break $\U(1)_r$ symmetry. The theory in the IR is in general in a Coulomb phase, $\U(1)$ massless gauge fields and massive hypermultiplets. We can try to use the Cartan generator of $\SU(2)_R$ to connect the IR and the UV theories on the generic locus of the Coulomb branch. This falls into the first scenario detailed above. The index computed this way is rather trivial and only captures the number of the $\U(1)$ gauge fields in the IR, the dimension of the Coulomb branch. 

However, on special loci of the Coulomb branch, scaling various Coulomb and mass parameters together, one can obtain an SCFT in the IR with an emergent $\U(1)_r$ R-symmetry. 
 For example, $\E_n$ Minahan-Nemeschansky theories are related between themselves and also related to ${\cal N}=2$ $\SU(2)$ $N_f=4$ SQCD by such Coulomb branch/mass deformations. The full indices of these theories fall into the second scenario above as they a~priori are unrelated.\footnote{ 
This logic breaks for a very special type of the ${\cal N}=2$ index, the Schur index. We discuss this in Appendix \ref{app:SchurWithMasses}. }

Surprisingly, it was observed in \cite{Deb:2025ypl} however that the non-relativistic limit of the ${\cal N}=2$ indices of theories related by flows breaking $\U(1)_r$ symmetry, at least in some cases, can be related to each other. Let us construct the non-relativistic limit using the definition of the full index \eqref{eq:FullIndex}. We parametrize,
\be
p=e^{-\epsilon},\qquad \frac{p q}t=p^\alpha\,,
\ee and will eventually take the limit $\epsilon\to 0$.
It is also convenient to define \cite{Gadde:2011uv},
\be
p=\sigma\tau,\qquad q=\rho\tau,\qquad t=\tau^2\,,
\ee 
and then the full index becomes,
\be
{\cal I}_{q,\alpha,\epsilon}=\Tr_{{\mathbb S}^3}\,(-1)^F\,
\rho^{2(R+j_2-j_1)}\,e^{-\epsilon(\Delta+j_1-j_2-2R)\,\alpha}\,
e^{-\epsilon(2j_2+2R)\,(1-\alpha)}\, e^{-\beta'(\Delta-2j_2-2R+r)}\,.
\ee Using the notations of \cite{Gadde:2011uv} this index counts states annihilated by  supercharge $\widetilde {\cal Q}_{1\dot{-}}$. We also notice the following relations,
\be
&&2\{{\cal Q}_{1+},\,\left({\cal Q}_{1+}\right)^\dagger\}=\Delta+2j_1-2R-r\equiv \delta\,,\qquad 
2\{\widetilde {\cal Q}_{1\dot{-}},\,\left(\widetilde{\cal Q}_{1\dot{-}}\right)^\dagger\}=\Delta-2j_2-2R+r\equiv \tilde \delta_1\,,\nonumber\\
&&2\{\widetilde {\cal Q}_{2\dot{+}},\,\left(\widetilde{\cal Q}_{2\dot{+}}\right)^\dagger\}=\Delta+2j_2+2R+r\equiv\tilde \delta_2\,,
\ee using which the index is written as,
\be
{\cal I}_{q,\alpha,\epsilon}=\Tr_{{\mathbb S}^3}\,(-1)^F\,
\rho^{2(R+j_2-j_1)}\,e^{-\epsilon\frac{\tilde\delta_1+\delta}2\,\alpha}\,
e^{-\epsilon\frac{\tilde\delta_2-\tilde\delta_1}2\,(1-\alpha)}\, e^{-\beta'\tilde\delta_1}\,.
\ee In the case of $\alpha=1$ the index counts actually states annihilated by two supercharges, ${\cal Q}_{1+}$ and $\widetilde {\cal Q}_{1\dot{-}}$, and is equal to the Schur index. In the case of $\alpha=0$ the index counts  states also annihilated by two supercharges, $\widetilde {\cal Q}_{2\dot{+}}$ and $\widetilde {\cal Q}_{1\dot{-}}$, and is equal to the massive/Coulomb index. In both of these cases it is manifestly independent of $\epsilon$ (and thus $p$) formally. 
Note that in principle in both of the special values of $\alpha$ above we can take $\beta'=0$ and then the index will be defined without using $\U(1)_r$ symmetry and thus well defined also for non-conformal theories (with the index defined to be computed with respect to some linear combination of the two preserved supercharges). See appendix \ref{app:SchurWithMasses} for more details and subtleties of the Schur index. For general value of $\alpha$ the index is well defined only for the conformal theory which possesses $\U(1)_r$.\footnote{See appendix \ref{app:multipletssection} for a discussion of contribution of superconformal multiplets in the ${\cal N}=2$ non-relativistic limit.}  

It was observed in \cite{Deb:2025ypl} that in some cases when two ${\cal N}=2$ conformal theories are related by vev/mass deformations their non-relativistic indices, \be {\cal I}_{q,\alpha}=\lim_{\epsilon\to0}{\cal I}_{q,\alpha,\epsilon}\,,\ee are related as,
\be \label{eq:GeneralizedSchurRelations}
{\cal I}_{q,\alpha_1}^{(1)} = {\cal I}_{q,\alpha_2=c_{12} \alpha_1}^{(2)}\,,
\ee for some value of constant $c_{12}$ depending on the two SCFTs.
We note that the computation of the two limits, $\Tr_{{\mathbb S}^3}$ as a partition function and $\epsilon\to 0$, do not in general commute. This rather surprising connection between indices is our main motivation to study further the non-relativistic limits of indices of various theories. 

Let us mention here two additional interesting properties that the generalized Schur limit of the ${\cal N}=2$ SCFTs has. The full supersymmetric index is known to possess certain $\text{SL}(3,{\mathbb Z})$ transformation properties related to anomalies \cite{Spiridonov:2011hf,Aharony:2013dha,Gadde:2020bov,MR1800253,MR2101221}.\footnote{See also \cite{DiPietro:2014bca,Bobev:2015kza,Shaghoulian:2016gol} for related works. Modularity(-like) properties also 
appear in three dimensional partition functions. See {\it e.g.} \cite{Aharony:2017adm,Cheng:2018vpl,Cheng:2024vou,ArabiArdehali:2026ddr}.}
This $\text{SL}(3,{\mathbb Z})$ property degenerates to $\text{SL}(2,{\mathbb Z})$ for the Schur index \cite{Razamat:2012uv}. Some of the modularity properties of the Schur index can be attributed to an underlying vertex operator algebra structure of any 4d $\mathcal{N}=2$ SCFT \cite{Beem:2013sza}. In particular, the Schur index solves a  modular linear differential equation (MLDE) \cite{Beem:2017ooy}. It has been conjectured recently that the generalized Schur limit also solves an MLDE  with coefficients depending on the parameter $\alpha$ \cite{Deb:2025ypl,Deb:2025ddc}. The conjecture was proven in a variety of cases in \cite{Chandra:2025qpv}.

The generalized Schur limit also indicates an interesting interplay between the Higgs and Coulomb branch, which are a priori two distinct branches of vacua. The Schur index counts Schur operators, which are a superset of the Higgs operators. Thus, the generalized Schur limit is naturally  related to the Higgs branch. As already mentioned above, in certain cases, the generalized Schur index captures the Schur index of theories related by renormalization group flows triggered by Coulomb branch vacuum expectation values. This already hints that this limit may have relations to the Coulomb branch as well. 
It was also shown in \cite{Deb:2025ddc} that for certain Argyres-Douglas theories, the generalized Schur limit for certain values of the parameter $\alpha$ equals the trace of higher powers of quantum monodromy $\mathcal{M}(q)$ (See also \cite{Cecotti:2015lab}, which studies a closely related limit and relates to powers of quantum monodromy). This generalizes a remarkable conjecture by \cite{Cordova:2015nma}, where it was shown that the Schur index equals the trace of the inverse of the quantum monodromy.\footnote{The inverse of the quantum monodromy is called the Kontsevich-Soibelman operator $\mathcal{O}(q)=\mathcal{M}(q)^{-1}$.} 

We will extend our study of the non-relativistic limit of supersymmetric indices in this paper in several ways. First, we will consider the relation of this limit to the non-relativistic limit of the Ruijsenaars–Schneider model. In particular we expect that the indices of class ${\cal S}$ theories in the generalized Schur limit to be expressible naturally using eigenfunctions of the elliptic Calogero-Moser model. 
These eigenfunctions are elliptic generalizations of the Jack polynomials. We will test this statement mainly for the $A_1$ class ${\cal S}$ case where the elliptic Calogero-Moser model is equivalent to the Lam\`{e} differential equation. In particular we will derive the eigenfunctions using a combination of two methods: 
diagonalizing the index \cite{Gadde:2011uv} 
and utilizing instanton partition functions in the presence of defects in five and four dimensions \cite{Hatsuda:2018lnv,Bullimore:2014upa,Nekrasov:2009rc,Kanno:2011fw,Kim:2024mnp}.\footnote{Let us mention that the groundstate of the integrable model can be rather easily obtained  using the limit of large compactifications \cite{Nazzal:2023wtw}.} This will be discussed in section \ref{sec:GeneralizedSchur}.

As an  example of an application of these results, in section \ref{sec:A2generalizedSchur} we will check an instance of the equality  \eqref{eq:GeneralizedSchurRelations} connecting the generalized Schur index of the rank one ${\mathfrak e}_6$ Minahan-Nemeschansky theory and the $\SU(2)$ ${\cal N}=2$ conformal SQCD using $A_1$ and $A_2$ elliptic Jack functions. In doing so, we provide a surprising identity between sums of the eigenfunction sets associated with two different root systems.

We will also extend the notion of the generalized Schur index to classes of ${\cal N}=1$ theories. First, in section \ref{sec:N1Schur} we will argue that even for ${\cal N}=1$ class ${\cal S}$ \cite{Benini:2009mz,Bah:2012dg,Agarwal:2015vla} the generalized Schur index is well defined and interesting for a  range of compactifications. The Schur limit of the index depends on special properties of ${\cal N}=2$ supersymmetry, but viewing it as a special case of the non-relativistic limit can be extended also to ${\cal N}=1$ compactifications of the $(2,0)$ SCFT.\footnote{This is to be contrasted with the mixed Schur limit discussed in \cite{Beem:2012yn} where the index depends on both $p$ and $q$.}

We will then further generalize this statement in section \ref{sec:Estring}. We will discuss the non-relativistic limit of indices of ${\cal N}=1$ theories obtained by compactifications of the $6d$ $(1,0)$ rank $Q$ E-string theory. It has been shown in \cite{Nazzal:2018brc} that the ${\cal N}=1$ indices of compactifications of the rank one E-string theory should be expressible in terms of eigenfunctions of the $BC_1$ van~Diejen model. This statement has a straightforward generalization to rank $Q$ E-string and the $BC_Q$ van~Diejen model. On the other hand the non-relativistic limit of the van~Diejen model is well studied and is given by the Inozemtsev model \cite{MR990577}. The Inozemtsev model can be also viewed as a generalization of the elliptic Calogero-Moser model with a single coupling $\alpha$ to a system with five (or four in case of $Q=1$) couplings. We will discuss this limit at the level of indices of theories obtained in the compactifications. In particular, yet again we will obtain various relations between theories that can be connected to each other by deformations. These deformations  include vev deformations in six dimensions and subsequent compactifications. See {\it e.g.} \cite{Razamat:2019mdt}. A special general feature of indices in non-relativistic limits that we consider is that these are expressible for theories with known Lagrangians as integrals of elliptic and modular functions, {\it i.e.} particular products of theta functions.\footnote{Another way to reduce the full ${\cal N}=1$ index to integrals over combinations of theta functions was discussed in \cite{Razamat:2020gcc}. It was observed that for non-chiral Lagrangian theories with quantized spectrum of R-charges setting $p$ to some power of $q$ (determined by the quantization of the R-charges), the elliptic Gamma functions appearing in the computation of the index reduce to theta functions. } 

\ 

\noindent We will finally summarize and discuss our results and possible further studies in section \ref{sec:summaryanddiscussion}.

    \section{Generalized Schur index for $A_1$ class ${\cal S}$ and the Calogero-Moser model }\label{sec:GeneralizedSchur}

	We begin our discussion of the non-relativistic limit with probably the simplest class of theories, the ${\cal N}=2$ $A_1$ class ${\cal S}$. All the theories in this class have an explicit ${\cal N}=2$ Lagrangian description. It is also expected that all theories in this class have their full indices expandable in terms of an orthonormal set of joint eigenfunctions of the following difference operators,
    \be\label{eq:A1RSmodel}
	&&{\cal D}_z\cdot f(z) =\frac{\theta_p(\frac{t}q z^{-2})}{\theta_p(z^{2})}  \,f(q^{\frac12}z)+
	\frac{\theta_p(\frac{t}q z^{2})}{\theta_p(z^{-2})}  \,f(q^{-\frac12}z)\,,\\
	&&\widetilde{\cal D}_z\cdot f(z) =\frac{\theta_q(\frac{t}p z^{-2})}{\theta_q(z^{2})}  \,f(p^{\frac12}z)+
	\frac{\theta_q(\frac{t}p z^{2})}{\theta_q(z^{-2})}  \,f(p^{-\frac12}z)\,.\nonumber
	\ee These are the $A_1$ RS operators~\cite{ruijsenaars1986new,ruijsenaars1987complete,ruijsenaars2004elliptic}. We refer the reader to appendix \ref{app:definitions} for definitions of various special functions and technical notations. Moreover, we are after eigenfunctions of these operators which are orthogonal under the Haar/vector multiplet measure,
    \be
    (q;q)(p;p)\oint\frac{dz}{4\pi i z}\frac{\Gamma_e(\frac{pq}t z^{\pm 2})}{\Gamma_e(z^{\pm2})}\psi_n(z)\psi_m(z)=\delta_{nm}\,.
    \ee For special limits such eigenfunctions are well known. For example,
    \be
    &\text{Schur: }(t=q)\;\;\;\; & \,\psi_n(z)\propto\frac{\chi^{(Schur)}_n(z)}{(qz^{\pm2};q)}\,,\\
    &\text{Macdonald: } (p=0)\;\;\;\; &\,\psi_n(z)\propto\frac{\chi^{(Mac)}_n(z;q,t)}{(t z^{\pm2};q)}\,,\nonumber\\
    &\text{HL: }(p=q=0)\;\;\;\;  &\,\psi_n(z) \propto\frac{\chi^{(HL)}_n(z;t)}{(1-t z^2)(1-t z^{-2})}\,.\nonumber
    \ee 
    Here, $\chi^{(Schur)}_n(z)$,  $\chi^{(Mac)}_n(z;q,t)$, and $\chi^{(HL)}_n(z;t)$ are the Schur, Macdonald polynomials, and Hall-Littlewood polynomials, respectively. We are after understanding the solutions to these equations in the non-relativistic limit. As we shall see now, one equation simplifies significantly in this regime, while the other reduces to a second-order differential equation. 
	
    \subsection{The non-relativistic limit of difference operators}

    Let us review explicitly the non-relativistic limit of the elliptic RS model. See for example \cite{ruijsenaars2000relativistic,ruijsenaars1999generalized,Hallnas:2024jhj}.
	  We consider the limit of the two operators \eqref{eq:A1RSmodel} taking $qp/t=p^\alpha$ and sending $p\to 1$:  we will parametrize $p=e^{-\epsilon}$ and send $\epsilon\to0$.\footnote{Note that unlike the $3d$ limit where all fugacities are sent to $1$, here one of the fugacities, $q$, is kept generic. For various aspects of the $3d$ limit of the index the readers can consult {\it e.g.} \cite{Dolan:2011rp,Gadde:2011ia,Niarchos:2012ah,Aharony:2013dha,ArabiArdehali:2015ybk,Cassani:2021fyv,ArabiArdehali:2021nsx}.} 
	
	For the first ($q$-shift) operator, we take the limit as
	\be
	\frac{\theta_p(\frac{t}q z^{-2})}{\theta_p(z^{2})}\;\;\to\;\; \lim_{p\to 1} \frac{\theta_p(p^{1-\alpha} z^{-2})}{\theta_p(z^{2})}=(-1)^\alpha z^{-2\alpha}\,,
	\ee and thus after multiplying by $(-1)^\alpha$, we get
	\be
	{\cal D}_z\cdot f(z)\;\;\to\;\; {\mathfrak D}_z\cdot f(z) =z^{-2\alpha} \,f(q^{\frac12}z)+z^{2\alpha} \,f(q^{-\frac12}z)\,.
	\ee 
    
	For the second ($p$-shift) operator, we compute in the limit,
	\be
	&&\frac{4(\ln\,q)^2}{\epsilon^2}\left(\widetilde{\cal D}_z-2-\alpha \epsilon\right)\cdot f(z)\;\to\; 
	z\frac{d}{dz}z\frac{d}{dz}\,f(z)+\alpha\left(z\frac{d}{dz}\ln\Delta(z)\right)\, z\frac{d}{dz}\,f(z)+\nonumber\\
	&&\frac12 \alpha^2 \left(2+\frac12\left(z\frac{d}{dz}\ln\Delta(z)\right)^2+z\frac{d}{dz}z\frac{d}{dz}\ln\Delta(z)\right)\,f(z)\equiv  \tilde {\mathfrak D}_z\cdot f(z)\,,
	\ee
    where we have defined the function $\Delta$ as
	\be
	\Delta(z)=\left[(1-z^{\pm2})\prod_{m=1}^\infty(1-z^{\pm2}q^m)^2\right]=\theta(z^{\pm2};q)\,.
	\ee Let us define $z=e^x$ and $f=e^{-\frac{\alpha}2\ln\Delta}\chi(x)$ and write,
	\be
	&&e^{\frac{\alpha}2\ln\Delta}\tilde {\mathfrak D}_z\cdot f(z)=e^{\frac{\alpha}2\ln\Delta}\tilde {\mathfrak D}_ze^{-\frac{\alpha}2\ln\Delta}\cdot \chi(z)=
	\chi''+\frac12\alpha(\alpha-1)(\ln\Delta)''\chi+\alpha^2\chi\,.
	\ee 
    Here {\it prime}s denote derivatives with respect to $x$.
     Our functions are joint eigenfunctions of the difference operator ${\mathfrak D}_z$ and the differential operator $\tilde {\mathfrak D}_z$.\footnote{In what follows we will derive expressions for eigenfunctions as a series in $q$. As such it will be easy to verify the eigenfunction property under the operator $\tilde {\mathfrak D}_z$ but somewhat technically harder under ${\cal D}_z$, as it mixes different orders of $q$.}
     
	The operator  $\tilde {\mathfrak D}_z$ is precisely the Lam\'{e} differential operator.
    To show the relation to Lam\'e equation, let us note that Weierstrass $\wp(x,\tau)$ can be expressed as follows,
	\be
		&& \wp(\xi,\tau)=-\frac{\partial^2}{\partial \xi^2}\log\theta_1(z|\tau)-\frac{\pi^2}{3}E_2(\tau)~,\qquad z=e^{\pi i \xi}\,,\qquad q=e^{2\pi i \tau}\,,\\
		&& \theta_1(z|\tau)=-iz^{\frac{1}{2}}q^{\frac{1}{8}}(q;q)(z q;q)(z^{-1};q)~,\qquad 
		\theta_1(z|\tau)^2=q^{\frac{1}{4}}(q;q)^2\Delta(z^{\frac{1}{2}})\,.\nonumber 
	\ee Here $E_2(\tau)$ is the quasi-modular Eisenstein series. 
	Therefore, $e^{\frac{\alpha}{2}\log\Delta}\tilde {\mathfrak D}_z e^{-\frac{\alpha}{2}\log\Delta}$ is proportional to the Lam\'e differential operator \cite{whittaker2020course} up to $z$ independent terms,\footnote{See also \cite{ruijsenaars2000relativistic, ruijsenaars1999generalized, ruijsenaars2003relativistic,ruijsenaars1999relativistic,maier2008lame,Langmann:2004sj,Atai:2016mir}.}
    \begin{equation}
		H_{\text{Lam\'e}}=-\frac{d^2}{d\xi^2}+g(g-1)\wp(\xi,\tau)~,~g\in \mathbb{R}~,
	\end{equation} with $g=\alpha$. The parameter $g$ is the value of the coupling. Sending $g$ to zero and to one the potential term vanishes. However, these two cases are physically different since we seek eigenfunctions that are orthonormal with respect to the $\alpha$-th power of the Haar/vector multiplet measure.\footnote{As the eigenfunctions under the vector multiplet measure are Weyl symmetric functions, normalizing the Schur eigenfunctions by the square root of the Haar measure yields Weyl-antisymmetric functions that are orthogonal under the trivial measure. In this sense, $g=1$ is a free fermion case and $g=0$ a free boson case. The $A_{N-1}$ integrable models describe systems with $N$ particles, where Weyl symmetry corresponds to the permutation symmetry of the particles. Thus, literally for $N=2$ in the Schur case, the eigenfunctions are antisymmetric and naturally describing fermions. }  The $g=1$ is related to the Schur limit while $g=0$ to the mass/Coulomb limit.
    We note that the further specialization $q=0$ gives the trigonometric Calogero-Moser model eigenfunctions of which are Jack polynomials that depend on the coupling $g$. For the $A_1$ case these are also known as the Gegenbauer polynomials. Although the indices become trivial in the $q\to 0$, it is still useful to organize the $q$ expansion of the eigenfunctions for the Lam\`{e} equation in terms of Jack polynomials. 
    We will often refer to the eigenfunctions of the non-relativistic elliptic Calogero-Moser model as {\it elliptic Jack functions}.

   \
    
	\subsection{Expansion of the $A_1$ trinions in elliptic Jack functions}

    The usual utility of solving for the eigenfunctions of \eqref{eq:A1RSmodel} is that the index of the ${\cal N}=2$ $A_1$ class ${\cal S}$ theory corresponding to a compactification on genus ${\mathfrak g}$ surface with ${\mathfrak s}$ punctures can be written as \cite{Gadde:2011uv,Gaiotto:2012xa},
       \begin{equation}\label{eq:generalsurfaceindexA1}
		\mathcal{I}_{\mathfrak{g},\mathfrak{s}}(\left\{z_I\right\})=\sum_{\lambda} \left(C_{\lambda}\right)^{2\mathfrak{g}+\mathfrak{s}-2} \prod_{I=1}^\mathfrak{s}\psi_{\lambda}(z_I)~.
	\end{equation} For the $A_1$ case at hand all the theories have explicit ${\cal N}=2$ Lagrangian descriptions, and thus their indices have an alternative  definition in terms of certain matrix model integrals.

	In this section, we consider two different expressions for generalized Schur  index of the compactification on three punctured sphere, (${\mathfrak g}=0, {\mathfrak s}=3$). On one hand, it is known that this theory is a trifundamental half-hypermultiplet of three $\SU(2)$ symmetries associated to the three punctures. The generalized Schur index can be directly evaluated via the Lagrangian description as
    \be \label{eq:freetrinionlagrangian}
    {\cal I}_{0,3}(z_1,z_2,z_3)=\frac1{(q^{\frac12}z_1^{\pm1}z_2^{\pm1}z_3^{\pm1};q)^\alpha}\,.
    \ee
    On the other hand, we should be able to write it as an expansion in eigenfunctions,
	\begin{equation}\label{eq:freetrinioneigenfunctions}
		\mathcal{I}_{0,3}(z_1,z_2,z_3)=\sum_\lambda C_{\lambda}\psi_\lambda(z_1;q,\alpha)\psi_\lambda(z_2;q,\alpha)\psi_\lambda(z_3;q,\alpha)\,.
	\end{equation}
    These two expressions must coincide. Although there are various methods to verify this, as we will discuss soon, we can actually exploit this equality to determine the eigenfunctions. 
    
    Let us first take the following ansatz for the eigenfunctions,
	\begin{equation}
		\psi_\lambda(z;q,\alpha)=\frac{\mathcal{F}_\lambda(q)}{(q\,z^{\pm2};q)^{\alpha}}\sum_{\lambda'=0}^\infty c_{\lambda\lambda'}(\alpha)J^{(\alpha)}_{\lambda'}(z) q^{\frac{|\lambda-\lambda'|}{2}}\,.
	\end{equation}
	Here $J_\lambda(z)$ denotes the $A_1$ Jack polynomials which are the well-known eigenfunctions of the Lam\'{e} equation in the limit $q\rightarrow 0$.
    The first few Jack polynomials take the following form.
	\begin{align}
		J^{(\alpha)}_0(z)&=1~,\nonumber\\
		J^{(\alpha)}_1(z)&=z+\frac{1}{z}~,\\
		J^{(\alpha)}_2(z)&=z^2+\frac{1}{z^2}+\frac{2 \alpha }{\alpha +1}~,\nonumber\\
		J^{(\alpha)}_3(z)&=z^3+\frac{1}{z^3}+\frac{3 \alpha }{\alpha +2}\left(z+\frac{1}{z}\right)\,, \cdots ~.\nonumber 
	\end{align}
    These polynomials are normalized to $|J_n^{(\alpha)}|^2=\frac{2 n! \Gamma (n+2 \alpha )}{(\alpha +n) \Gamma (n+\alpha )^2}$.
	These expressions reduce to Schur polynomials in the limit $\alpha\to 1$. With this, one can solve for the coefficients $c_{\lambda\lambda'}(\alpha)$ by comparing  the two expressions \eqref{eq:freetrinionlagrangian} and \eqref{eq:freetrinioneigenfunctions} order by order in $q$. This leads to the following expressions for the lowest-lying orthonormal eigenfunctions, 
       \begin{align}
	&\psi_0(z)=\frac{(qz^{\pm 2};q)^{-\alpha}}{\sqrt{\frac{2 \Gamma (2 \alpha )}{\alpha  \Gamma (\alpha )^2}+\frac{4 \left(\alpha ^3-3 \alpha +2\right) \alpha ^2 q^2 \Gamma (2 \alpha +2)}{(\alpha +1)^2 \Gamma (\alpha +3)^2}}}\left(J^{\alpha }_0(z)+\frac{(\alpha -1) \alpha  q J^{\alpha }_2(z)}{\alpha +1}+\right.\nonumber\\
    & \;\;\left.\frac{(\alpha\! -\!1) \alpha  q^2 \left(\left(\alpha ^3\!+\!4 \alpha ^2\!+\!5 \alpha \!+\!6\right) (\alpha \!+\!1)^2 J^{\alpha }_4(z)+2\! \left(\alpha ^4\!-\!17 \alpha ^3\!-\!39 \alpha ^2\!+\!\alpha \!+\!6\right) J^{\alpha }_2(z)\right)}{2 (\alpha +1)^3 (\alpha +2) (\alpha +3)}\!+\!{\cal O}(q^3)\!\right)\,,\nonumber\\
	&\psi_1(z)=\frac{(qz^{\pm 2};q)^{-\alpha}}{\sqrt{\frac{2 \Gamma (2 \alpha +1)}{(\alpha +1) \Gamma (\alpha +1)^2}}}\left(J^{\alpha }_1(z)+\frac{(\alpha -1) \alpha  q J^{\alpha }_3(z)}{\alpha +2}+{\cal O}\left(q^2\right)\right)\,,\\
	&\psi_2(z)=\frac{(qz^{\pm 2};q)^{-\alpha}}{\sqrt{\frac{4 (\alpha +2) \Gamma (2 \alpha +2)}{\Gamma (\alpha +3)^2}}}\left(J^{\alpha }_2(z)+q \left(\frac{(\alpha -1) \alpha  J^{\alpha }_4(z)}{\alpha +3}-\frac{4 \alpha  \left(2 \alpha ^2-\alpha -1\right) J^{\alpha }_0(z)}{(\alpha +1)^3 (\alpha +2)}\right)+{\cal O}\left(q^2\right)\right)\,,\nonumber\\
	&\psi_3(z)=\frac{(qz^{\pm 2};q)^{-\alpha}}{\sqrt{\frac{12 (\alpha +3) \Gamma (2 \alpha +3)}{\Gamma (\alpha +4)^2}}}J^{\alpha }_3(z)+{\cal O}\left(q^1\right)\,,\qquad 
	\psi_4(z)=\frac{(qz^{\pm 2};q)^{-\alpha}}{\sqrt{\frac{48 (\alpha +4) \Gamma (2 \alpha +4)}{\Gamma (\alpha +5)^2}}}J^{\alpha }_4(z)+{\cal O}\left(q^1\right)\,.\nonumber
\end{align}
The structure constants are given by,

    	\begin{align}
		C_{0}&=\left(\frac{2 \Gamma (2 \alpha )}{\alpha  \Gamma (\alpha )^2}+\frac{4 \left(\alpha ^3-3 \alpha +2\right) \alpha ^2 q^2 \Gamma (2 \alpha +2)}{(\alpha +1)^2 \Gamma (\alpha +3)^2}\right)^{\frac{3}{2}}\left(1+\frac{q \left(10 \alpha ^2+2 \alpha ^3\right)}{(1+\alpha )^2}+\right.\nonumber\\
        &\;\; \left.\frac{q^2 \left(78 \alpha ^2+125 \alpha ^3+199 \alpha ^4+141 \alpha ^5+31 \alpha ^6+2 \alpha
			^7\right)}{(1+\alpha )^3 (2+\alpha )^2}+{\cal O}(q^3)\right)\,,\nonumber\\
		C_{1}&=\left(\frac{2 \Gamma (2 \alpha +1)}{(\alpha +1) \Gamma (\alpha +1)^2}\right)^{\frac{3}{2}}\left(\sqrt{q} \alpha +\frac{q^{3/2} \alpha  \left(4-6 \alpha +18 \alpha ^2+2 \alpha ^3\right)}{(2+\alpha )^2}+{\cal O}(q^{5/2})\right)\,,\\
		C_{2}&=\left(\frac{4 (\alpha +2) \Gamma (2 \alpha +2)}{\Gamma (\alpha +3)^2}\right)^{\frac{3}{2}}\left(\frac{1}{2} q \alpha  (1+\alpha )+\frac{q^2 \alpha  \left(24 \alpha +10 \alpha ^2+28 \alpha ^3+2 \alpha ^4\right)}{2 (3+\alpha )^2}+{\cal O}(q^3)\right)\,,\nonumber
        \end{align}
        \begin{align}
        C_{3}&=\left(\frac{12 (\alpha +3) \Gamma (2 \alpha +3)}{\Gamma (\alpha +4)^2}\right)^{\frac{3}{2}}\left(\frac{1}{6} \alpha  \left(\alpha ^2+3 \alpha +2\right) q^{3/2}+{\cal O}(q^{5/2})\right)\,,\nonumber\\
        C_{4}&=\left(\frac{48 (\alpha +4) \Gamma (2 \alpha +4)}{\Gamma (\alpha +5)^2}\right)^{\frac{3}{2}}\left(\frac{1}{24} \alpha  \left(\alpha ^3+6 \alpha ^2+11 \alpha +6\right) q^2+{\cal O}(q^3)\right)\,.\nonumber
	\end{align}
    Using the wavefunctions derived above we can write the generalized Schur partition function (the non-relativistic limit of the full index) of a theory associated to a Riemann surface of genus $\mathfrak{g}$ with $\mathfrak{s}$ puncture, with the flavor fugacities turned off, as in \eqref{eq:generalsurfaceindexA1},\footnote{In principle one can turn on fugacities for global symmetries on. However, we refrain from doing so as we want to relate the non relativistic indices of theories related by RG flows and having  different flavor symmetries to each other. Let us also mention a subtlety defining the non-relativistic limit of a theory with a Coulomb branch discussed in \cite{Deb:2025ypl}. In the non-relativistic limit (with $\alpha\neq 1$), the indices of gauge theories exhibit pole divergences associated with Coulomb operators, $\Tr\Phi^n$, order of which is the dimension of the Coulomb branch. Technically, these divergences come from $\Gamma_e(\frac{p q}t)$ factors in the index. Thus, when one defines 
    the index of an interacting SCFT in the non-relativistic limit we should take the leading Laurent coefficient in this limit and choose a convenient normalization (for example normalize so that the index will start with $1$). See \cite{Deb:2025ypl} for details.}
    \begin{equation}
        \mathcal{I}_{\mathfrak{g},\mathfrak{s}}(q)=(q;q)^{2 (3 \mathfrak{g}+\mathfrak{s}-3)}\sum_\lambda \left(C_\lambda\right)^{2\mathfrak{g}+\mathfrak{s}-2}\psi_\lambda(1)^{\mathfrak s}\,.
    \end{equation}
    
    As an example, consider the four-punctured sphere, \textit{i.e.,} $\mathcal{N}=2$ conformal SQCD. After normalizing the index such that the leading-order $q^0$ term is one, it takes the following form:
    \begin{equation}\label{eq:d4result}
        \mathcal{I}_{0,4}(q)=1+\frac{2  (-1+5 \alpha ) (1+6 \alpha )}{1+\alpha }q+\frac{-2-3 \alpha +29 \alpha ^2+150 \alpha ^3+1800 \alpha ^4}{(1+\alpha ) (2+\alpha )}q^2+{\cal O}(q^3)~.
    \end{equation}
    This matches the expression obtained using a direct evaluation of the contour integral or by solving an MLDE, as was done in \cite{Deb:2025ddc}. In particular, by specializing $\alpha=1,\,2\,,3\,,5,\,\frac12, \frac13, \frac15$, we obtain the expansions \footnote{Although here we provide examples for positive values of $\alpha$, the expression \eqref{eq:d4result} can be used to analytically continue to negative values of $\alpha$. The negative values of $\alpha$ were considered in \cite{Deb:2025ddc, Chandra:2025qpv}, where it was  shown to match in some cases higher powers of quantum monodromy traces considered in \cite{Kim:2024dxu}.
    The analytically continued index diverges  at negative integer values of $\alpha$. It would be interesting to understand the Physics of these divergences.}
    \be
    \begin{split}
   &1+28 q+329q^2+\cdots,\quad 1+78 q+ 2509q^2+
\cdots,\quad 1+133 q+ 7505 q^2+\cdots,\quad \\& 1+248 q+ 27249q^2+\cdots,\quad  1+8 q + 36q^2+\cdots,\quad  1+3 q + 9q^2+\cdots,\quad \\& 1+0q+q^2+\cdots \,,\;\;\;\;
\end{split}
    \ee
    which precisely match the Schur indices of ${\mathfrak d}_4$, ${\mathfrak e}_6$, ${\mathfrak e}_7$,  ${\mathfrak e}_8$ SCFTs, and the  ${\mathfrak a}_2$, ${\mathfrak a}_1$, ${\mathfrak a}_0$ AD models, respectively.
    
     We can also solve for the eigenfunctions by explicitly solving for the spectral problem of the Lam\`{e} equation. Moreover, 
     one can utilize the relation to $4d$ reductions of $5d$ instanton partition functions. As this relation is very useful we will discuss it next.

    \subsection{$A_1$ Elliptic Jack functions from instanton computations}\label{subsec:A1-instanton}

    In this subsection we present a complementary approach to the eigenfunctions of the elliptic Calogero-Moser model based on instanton calculus in 4d $\mathcal{N}=4$ $\SU(N)$ supersymmetric Yang-Mills theory in the presence of codimension-two and codimension-four defects. Our starting point is the elliptic $A_{N-1}$ Ruijsenaars-Schneider model and its realization in the 5d $\mathcal{N}=2$ $\SU(N)$ gauge theory with $1/2$-BPS defects. Taking the non-relativistic limit then leads to the elliptic Calogero-Moser model, as discussed above, and in this limit the corresponding wavefunctions are naturally identified with ramified instanton partition functions of the 4d $\mathcal{N}=4$ theory with a codimension-two defect.

We begin by recalling the 5d construction of the elliptic RS wavefunctions developed in \cite{Bullimore:2014awa}. Consider the 5d $\mathcal{N}=2$ $\SU(N)$ Yang-Mills theory on the Omega-deformed background ${\mathbb S}^1\times \mathbb{R}^4_{\epsilon_1,\epsilon_2}$, where $\epsilon_{1,2}$ are the Omega-deformation parameters. This theory admits a natural $1/2$-BPS codimension-two defect, namely the Gukov-Witten monodromy defect \cite{Gukov:2006jk,Gukov:2008sn}, which prescribes a monodromy for the gauge field around the defect locus. An equivalent description of this defect is obtained by coupling a 3d $\mathcal{N}=4$ quiver gauge theory called $T[SU(N)]$ theory to the 5d bulk theory \cite{Gaiotto:2008ak}, with the 3d theory supported on ${\mathbb S}^1\times \mathbb{R}^2_{\epsilon_1}$.

The connection to integrable models emerges as follows. In the Seiberg-Witten limit $\epsilon_{1,2}\rightarrow0$, the supersymmetric vacua of the coupled 3d/5d system are governed by the twisted chiral ring relations \cite{Gaiotto:2013bwa}, which coincide with the spectral curve of the elliptic $A_{N-1}$ RS integrable system. Turning on the parameter $\epsilon_1$ then quantizes these relations, with $\epsilon_1$ playing the role of the Planck constant. The resulting  twisted chiral ring relations take the form of the difference equations acting on a wavefunction $\Psi_{\rm RS}$ as \cite{Bullimore:2014awa}
\begin{align}
\sum_{\substack{I\subset \{1,\dots,N\}\\ |I|=r}}
\prod_{\substack{i\in I\\ j\notin I}}
\frac{\theta_{1}(\eta^{2}z_i/z_j;q)}{\theta_{1}(z_i/z_j;q)}
\prod_{i\in I} y_i\ \Psi_{\rm RS}(z_i)
=
W_r(\mu_1,\dots,\mu_N;q\ \Psi_{\rm RS}(z_i),
\qquad r=1,\dots,N\!-\!1,
\label{eq:RS-Hamiltonians}
\end{align}
where $z_i$ are the Fayet--Iliopoulos (FI) parameters of the 3d $T[\SU(N)]$ theory, and $y_i$ are shift operators satisfying $y_iz_j = p^{\delta_{ij}}z_jy_i$.  $q$ is the elliptic parameter related to the 5d gauge coupling, $\eta$ is the mass parameter for the adjoint hypermultiplet (written in the conventions of 5d $\mathcal{N}=1$ supersymmetry), and $W_r$ denotes the expectation value of Wilson loop in $r$-th antisymmetric representation, expressed in terms of the Coulomb branch parameters $\mu_i$.

The left-hand sides of \eqref{eq:RS-Hamiltonians} are precisely the $N-1$ independent, mutually commuting Hamiltonians of the elliptic $A_{N-1}$ RS model. From the perspective of 4d $A_{N-1}$-type Class $\mathcal{S}$ theories, these operators act on the degrees of freedom associated to maximal punctures to generate half-BPS surface defects \cite{Gaiotto:2012xa} whose expectation values are identified with the Wilson loops on the right-hand side. It is also worth noting that in the limit $q\to 0$, the 3d theory decouples from the 5d bulk, and the difference equations reduce to those of the trigonometric RS model.

The eigenfunctions of these difference operators are obtained from the partition function of the coupled 3d/5d system in the Nekrasov-Shatashvili (NS) limit $\epsilon_2\to 0$. This fact has a natural explanation from the 6d perspective underlying class $\mathcal{S}$ theories. Each eigenfunction $\psi_\lambda$ is associated to a puncture that arises from inserting a codimension-two defect in six dimensions. Upon circle compactification, the maximal puncture gives rise precisely to the 3d $T[\SU(N)]$ defect in the 5d $\mathcal{N}=2$ theory. This naturally explains the relation between the eigenfunction and the partition function of the 3d/5d coupled system \cite{Bullimore:2014awa,Bullimore:2014upa,Kim:2024mnp}, which thereby establishes the dictionary between eigenfunctions and 3d/5d partition functions \cite{Bullimore:2014awa,Bullimore:2014upa,Kim:2024mnp}.

More concretely, the partition function of the 3d/5d system on ${\mathbb S}^1\times \mathbb{R}^4$ can be computed by supersymmetric localization and ramified instanton counting \cite{Alday:2010vg,Kanno:2011fw}. At this stage, however, the resulting object should be regarded as a \emph{formal} eigenfunction of the elliptic RS Hamiltonians. It still depends on the Coulomb branch parameters $\mu_i$, and therefore is not yet directly comparable to the physical eigenfunctions discussed in the previous subsections. In particular, it is not $L^2$-normalizable. To promote it to a genuine normalizable wavefunction, one must impose an appropriate quantization condition on the Coulomb branch parameters \cite{Nekrasov:2009rc}. We will postpone this issue until after taking the non-relativistic limit. We also refer to \cite{Grassi:2014zfa,Marino:2016rsq,Hatsuda:2018lnv} for discussions of such quantization conditions and the associated non-perturbative completions of the 5d partition functions.

We now take the non-relativistic limit $\epsilon\rightarrow0$ with
\begin{align}
	p=e^{-\epsilon} \ , \quad \eta = e^{-\epsilon\alpha/2} \ , \quad \mu_i=e^{-\epsilon a_i} \ ,
\end{align}
which corresponds to shrinking the ${\mathbb S}^1$ radius proportional to $\epsilon$ to zero. Under this limit the 5d theory reduces to the 4d $\mathcal{N}=4$ $\SU(N)$ theory, with the codimension-two monodromy defect now described by a 2d reduction of the $T[\SU(N)]$ theory. Then the difference equation \eqref{eq:RS-Hamiltonians} for $r=1$ in this limit becomes a differential equation as
\begin{align}\label{eq:eCM-hamiltonian-AN}
	\left[-\frac{1}{2}\sum_{i=1}^N \frac{\partial^2}{\partial x_i^2}+\alpha(\alpha-1)\sum_{i<j}\wp(x_i-x_j|\tau)\right]\Psi(x_i,a_j,\alpha;q) = \mathcal{E} \Psi(x_i,a_j,\alpha;q)\ ,
\end{align}
with $z_i=e^{2\pi ix_i}$.
The operator on the left-hand side is precisely the quadratic Hamiltonian of the elliptic Calogero-Moser model. As we explain below in \eqref{eq:chiral-ring}, the eigenvalue $\mathcal{E}$ is related to the expectation value of the chiral ring operator obtained by reducing the 5d fundamental Wilson loop wrapping ${\mathbb S}^1$ to four dimensions.

The wavefunction (or eigenfunction) $\Psi$ of the above differential equation is given by the normalized partition function of the resulting 2d/4d coupled system in the NS limit $\epsilon_2\rightarrow0$:
\begin{align}\label{eq:normalized-defect-partition-function}
	\Psi(x_i,a_j,\alpha;q) &= \prod_{i<j}^N\theta_1(z_i/z_j;q)^\alpha\cdot D(x_i,a_j,\alpha;q) \ , \nonumber \\
	D(x_i,a_j,\alpha;q) &=\prod_{i=1}^Nz_i^{a_i}\cdot (z_1/z_{N})^\alpha \cdot \lim_{\epsilon_2\rightarrow0}\frac{Z^{2d/4d}_{\rm inst}(z_i,a_j,\alpha;q)}{Z^{4d}_{\rm inst}(a_j,\alpha;q)} \ .
\end{align}
\begin{align}\label{eq:normalized-defect-partition-function}
	\Psi(x_i,a_j,\alpha;q) &= \prod_{i<j}^N\theta_1(z_i/z_j;q)^\alpha\cdot D(x_i,a_j,\alpha;q) \ , \nonumber \\
	D(x_i,a_j,\alpha;q) &=\prod_{i=1}^Nz_i^{a_i}(z_i/z_{i+1})^\alpha \cdot \lim_{\epsilon_2\rightarrow0}\frac{Z^{2d/4d}_{\rm inst}(z_i,a_j,\alpha;q)}{Z^{4d}_{\rm inst}(a_j,\alpha;q)} \ .
\end{align}
The functions $Z^{2d/4d}_{\rm inst}$ and $Z^{4d}_{\rm inst}$ denote the 4d instanton partition functions with and without the codimension-two defect, respectively.
Here, $a_j$ are the (classical) Coulomb branch parameters of the $\SU(N)$ gauge theory, $\alpha$ is identified with the $\mathcal{N}=2^*$ mass parameter for the adjoint matter, $\tau$ with $q\equiv e^{2\pi i \tau}$ is the 4d gauge coupling (which simultaneously serves as the instanton counting parameter), and the $x_i$ are the FI parameters in the 2d $T[\SU(N)]$ theory with condition $\sum_ix_i=0$. The prefactor $\prod_{i<j}^N \theta(z_i/z_j;q)^\alpha$ in the first line is included to eliminate the first-derivative terms in the Hamiltonian in the non-relativistic limit, whereas the factor $\prod_{i}z_i^{a_i}(z_i/z_{i+1})^\alpha$ in the second line is the classical contribution from the 2d defect partition function.

Let us now briefly review how the ramified instanton partition function and the chiral observable vevs are computed. A codimension-two defect of Gukov-Witten type \cite{Gukov:2006jk} is specified by a monodromy for the gauge field around the defect locus $\mathbb{R}^2_{\epsilon_1}\subset\mathbb{R}^4$. This monodromy is labeled by a partition of $N$,
\begin{align}
	\rho=[n_1,n_2,\ldots,n_s]\, , \qquad  \qquad \sum_{a=1}^sn_a= N \ ,
\end{align}
which breaks the gauge group to the Levi subgroup
\begin{align}
	\text{SU}(N) \quad \rightarrow \quad  \text{S}[\U(n_1)\times\cdots\times \U(n_s)] \ .
\end{align}
In the absence of the defect, instanton sectors are labeled by a single topological charge $k$. The defect refines this into a collection of fractional instanton numbers
\begin{equation}
\{k_1,\cdots,k_s\}, \qquad \qquad k=\sum_{j=1}^s k_j ,
\end{equation}
one for each factor of the Levi subgroup. The corresponding instanton moduli space, called the \emph{ramified instanton moduli space}, admits a description via the chain-saw quiver construction introduced in \cite{Kanno:2011fw}, which generalizes the standard ADHM construction to accommodate the codimension-two defect.

The ramified instanton partition function is then evaluated by equivariant localization. The fixed points are labeled by collections of Young diagrams $\lambda_{j,\ell}$ as 
\begin{equation}
\vec\lambda=\{\lambda_{j,\ell}\}, \qquad
j=1,\cdots,s, \qquad \ell=1,\cdots,n_j ,
\end{equation}
and the partition function takes the form
\begin{equation}
Z_\rho = \sum_{\vec\lambda}
q_1^{k_1(\vec\lambda)} \cdots q_s^{k_s(\vec\lambda)}
\, Z_{\rho,\vec\lambda}.
\label{eq:ramifiedZ-general}
\end{equation}
Here the instanton fugacities $q_i$ satisfy
\begin{equation}
q_1 \cdots q_s = q ,
\end{equation}
where $q=e^{2\pi i\tau}$ is the ordinary 4d gauge coupling. Each contribution $Z_{\rho,\vec\lambda}$ is computed by localization on the ramified instanton moduli space, or equivalently via the chain-saw quiver description of the 2d defect theory. There is also an equivalent orbifold description \cite{Kanno:2011fw}, in which the ordinary ADHM construction is placed on
\begin{equation}
\mathbb{C}_{\epsilon_1}\times (\mathbb{C}_{\epsilon_2}/\mathbb{Z}_\rho),
\end{equation}
with the orbifold action $\mathbb{Z}_\rho$ determined by the partition $\rho$. The fixed points of this orbifold action are again labeled by the same collections of Young diagrams $\vec\lambda$. This provides a geometric realization of ramified instanton counting for the 2d monodromy defect.

The wavefunction of the elliptic CM Hamiltonian in particular corresponds to the maximal monodromy $\rho=[1,1,\cdots,1]$, which fully breaks $\SU(N)$ to $\U(1)^{N-1}$. Concretely, the partition function appearing in \eqref{eq:normalized-defect-partition-function} is
\begin{align}
	Z^{2d/4d}_{\rm inst}(z_i;q) = Z_{\rho=[1,1,\cdots,1]}(q_i,q) \ .
\end{align}
The fractional instanton fugacities are related to the FI parameters of the 2d defect theory by
\begin{align}
	q_i = z_i/z_{i+1}=e^{x_i-x_{i+1}} \ \ {\rm for} \ \ i=1,2,\cdots,N-1 \,, \quad  q_{N} = (z_N/z_1)q= e^{x_N-x_1}q \ .
\end{align}

As a concrete illustration, for the 4d $\mathcal{N}=4$ $\SU(2)$ gauge theory the normalized defect partition function in the NS limit expands as
\begin{align}\label{eq:A1-partition-function}
	 \frac{Z^{2d/4d}_{\rm inst}}{Z^{4d}_{\rm inst}}=& 1 + q_1\frac{\alpha(2a\!+\!\alpha)}{2a\!+\!1}+q_1^2\frac{\alpha(1\!+\!\alpha)(2a\!+\!\alpha)(2a\!+\!\alpha\!+\!1)}{2(2a\!+\!1)(2a\!+\!2)} + \left(q_1\rightarrow q_2\,\&\,a\rightarrow-a\right) \\
	 &+q_1q_2\frac{\alpha\left(8\alpha a^3+4\alpha a^2-2(\alpha^3-4\alpha^2+6\alpha-2)a-3\alpha^3+8\alpha^2-8\alpha+2\right)}{(2a-1)(2a+1)^2}+\cdots \ ,\nonumber
\end{align}
where $a_1=-a_2=a$. As expected, this expression still depends on the continuous Coulomb parameter $a$, so it does not yet define a physical $L^2$-normalizable wavefunction. To obtain the physical solution, we must impose an appropriate quantization condition.

For 4d $\mathcal{N}=4$ $\SU(N)$ gauge theory, the quantization condition proposed in \cite{Nekrasov:2009rc} is
\begin{align}
	a_i - a_{i+1} = -\alpha - n_i \,, \quad n_i\in \mathbb{Z}_{\ge0} \,, \quad i=1,2,\cdots N-1 \ .
\end{align}
This agrees with the B-model quantization condition used for $L^2$-normalizable wavefunctions in the elliptic RS model in \cite{Hatsuda:2018lnv,Kim:2024mnp}. We claim that, after imposing this condition, the defect partition functions in \eqref{eq:normalized-defect-partition-function} become the physical eigenfunctions of the elliptic CM Hamiltonian.

For example, imposing the $\SU(2)$ quantization condition $2a=-\alpha-n$ in \eqref{eq:A1-partition-function}, we obtain the first few eigenfunctions for the elliptic $A_1$ CM model:
\begin{align}
	D_{n=0} &= 1+q\frac{2\alpha}{\alpha+1}\left(\alpha\chi_{\bf 3}(z)+2\alpha-1\right)+q^2\frac{\alpha^2(2\alpha+1)\chi_{\bf 5}(z)}{\alpha+2}  \\
    &+q^2\frac{\alpha^2(10\alpha^4\!+\!25\alpha^3\!+\!3\alpha^2\!+\!3\alpha\!+\!7)\chi_{\bf 3}(z)\!+\!10\alpha^6\!+\!15\alpha^5\!-\!33\alpha^4\!+\!3\alpha^3\!+\!11\alpha^2\!-\!6\alpha}{(\alpha+2)(\alpha+1)^3}+\mathcal{O}(q^3) \nonumber \\
	D_{n=1} &=\chi_{\bf 2}(z)+q\frac{\alpha(2\alpha+1)\chi_{\bf 4}(z)+(6\alpha^2-10\alpha+4)\chi_{\bf 2}(z)}{\alpha+2}+\mathcal{O}(q^2)  \nonumber 
\end{align}
\begin{align}
	D_{n=2} &=\chi_{\bf 3}(z)\!+\!\frac{2\alpha}{\alpha\!+\!1}\!+\!q\frac{2\alpha(\alpha\!+\!1)^3\chi_{\bf 5}(z)\!+\!4\alpha(2\alpha^2\!-\!\alpha\!-\!1)\chi_{\bf3}(z)\!+\!2\alpha(3\alpha^2\!-\!7\alpha^2\!+\!7\alpha\!+\!5)}{(\alpha+3)(\alpha+1)^2} \!+\!\mathcal{O}(q^2) \nonumber \\
	D_{n=3} &=\chi_{\bf 4}(z)+\frac{2(\alpha-1)}{\alpha+2}\chi_{\bf 2}(z)+q\frac{\alpha(2\alpha+3)\chi_{\bf 6}(z)}{\alpha+4}  \nonumber \\
	&\qquad  +q\frac{2\alpha(5\alpha^3+4\alpha^2-5\alpha-4)\chi_{\bf 4}(z)+\alpha(14\alpha^3-29\alpha^2+40\alpha+20)\chi_{\bf 2}(z)}{(\alpha+4)(\alpha+2)^2}+\mathcal{O}(q^2) \ ,\nonumber
\end{align}
where $\chi_{\bf r}(z)$ denotes the $\SU(2)$ character in the ${\bf r}$-dimensional representation with fugacities $z_i$. One can check that these are in perfect agreement with the wavefunctions $\psi_{\lambda=0,1,2,3}$ computed in the previous subsection, up to a $z_i$-independent prefactor.\footnote{In particular the $q\to 0$ expressions give the Jack polynomials.} 

To complete the picture, we note that the eigenvalues of the CM Hamiltonian also admit a direct description in terms of 4d partition functions. For the RS Hamiltonians, the eigenvalues are provided by the expectation values of the 5d Wilson loop in the fundamental representation of $\SU(N)$ \cite{Gaiotto:2014ina,Bullimore:2014awa}. Under dimensional reduction to 4d, the 5d Wilson loop wrapped around the circle reduces to the chiral ring operator associated to the second Casimir invariant of $\SU(N)$ in the $\mathcal{N}=4$ theory. This leads to relation
\begin{align}\label{eq:chiral-ring}
	\mathcal{E}\equiv \langle {\rm Tr}\, a^2\rangle - \frac{N(N\!-\!1)}{2}\alpha(\alpha\!-\!1)\frac{E_2(q)}{24}
	= \lim_{\epsilon\rightarrow0}\frac{W_{\rm fund}\!-\!N}{\epsilon^2} - \frac{N(N\!-\!1)}{2}\alpha(\alpha\!-\!1)\frac{E_2(q)}{24},
\end{align}
where $W_{\rm fund}$ denotes the expectation value of the 5d fundamental Wilson loop on the circle. The term proportional to $E_2(q)$ is included so that the Weierstrass $\wp(x,\tau)$ appearing in the CM Hamiltonian is written in its standard convention, which differs by an $E_2(q)$-dependent shift.

The 4d chiral ring vev $\langle {\rm Tr}\, a^2\rangle$ can be computed either via equivariant localization on the instanton moduli space in the presence of the universal bundle \cite{Gaiotto:2014ina,Bullimore:2014awa}, or through the qq-character formalism \cite{Nekrasov:2015wsu}. More precisely, both methods naturally yield the result for the $\U(N)$ theory, so one must subtract the $\U(1)$ contribution to obtain the $\SU(N)$ answer. The resulting vev is therefore
\begin{align}
	\langle {\rm Tr}\, a_{\SU(N)}^2\rangle
	&= \langle {\rm Tr}\, a_{\U(N)}^2\rangle - N\,\mathcal{E}_{\U(1)},
	\nonumber\\
	\mathcal{E}_{\U(1)}
	&= \lim_{\epsilon\rightarrow0}\frac{W_{\U(1)}-1}{\epsilon^2}
	= \alpha(\alpha-1)\left( \frac{E_2(q)}{24}-\frac{1}{24}\right)\, ,
\end{align}
where the $\U(1)$ factor is extracted from the 5d $\U(1)$ Wilson loop \cite{Bullimore:2014awa},
\begin{align}
	W_{\U(1)}=
	\frac{(q/\eta^2;q)_{\infty}(q\eta^2/p;q)_{\infty}}
	{(q;q)_{\infty}(q/p;q)_{\infty}} \ ,
\end{align}
in the 4d or non-relativistic limit.

For the $\SU(2)$ theory, the formal eigenvalue extracted from the expectation value of the chiral ring operator takes the form
\begin{align}
	&\mathcal{E}_{\SU(2)}
	=  \langle {\rm Tr}\, a_{\SU(2)}^2\rangle - \alpha(\alpha-1)\frac{E_2(q)}{24} \\
    \langle {\rm Tr}\, &a_{\SU(2)}^2\rangle=\,
	a^2
	+ q\frac{2\alpha^2(\alpha-1)^2}{4a^2-1}
	+ q^2\frac{192\alpha^2(\alpha-1)^2\left(2a^6-(\alpha^2-\alpha+2)a^4\right)}{2(a^2-1)(4a^2-1)^3} \nonumber
	\\
	& 
	+ q^2\frac{\alpha^2(\alpha\!-\!1)^2\left(4(5\alpha^4\!-\!10\alpha^3\!+\!29\alpha^2\!-\!24\alpha\!+\!30)a^2\!+\!7\alpha^4\!-\!14\alpha^3\!-\!5\alpha^2\!+\!12\alpha\!-\!12\right)}{2(a^2-1)(4a^2-1)^3}
	+ \mathcal{O}(q^3) \, .
	\nonumber
\end{align}
After imposing the quantization condition $2a=-\alpha-n$, one obtains the discrete spectrum of eigenvalues at level $n$. One can then verify that the resulting expression precisely reproduces the eigenvalues of the elliptic CM Hamiltonian in \eqref{eq:eCM-hamiltonian-AN}.

    \subsection{Schur limit of ${\cal N}=1$ class ${\cal S}$}\label{sec:N1Schur}
    Until now we have discussed the ${\cal N}=2$ class ${\cal S}$ compactifications. However, one can generalize this setup to compactifications of $A_1$ $(2,0)$ theory preserving only ${\cal N}=1$ supersymmetry \cite{Benini:2009mz,Bah:2012dg,Bah:2011je,Agarwal:2015vla}.\footnote{
    Yet another generalization is to consider punctures which preserve only ${\cal N}=1$ supersymmetry \cite{Xie:2013gma,Heckman:2016xdl,Apruzzi:2025znw}.
    We do not discuss it here.
    }
    One way to view the generalization is as follows. The R-symmetry in six dimensions is ${\frak{ usp}}(4)$. In terms of the $(1,0)$ six dimensional superconformal algebra we  can identify an ${\mathfrak {su}}(2)$ sub-algebra of this as the R-symmetry while the commutant ${\mathfrak {su}}(2)$ is global symmetry. Upon compactification to four dimension we need to choose a twist preserving supersymmetry.
    We can organize our thinking by first considering the R-symmetry twist preserving the ${\mathfrak {su}}(2)$ global symmetry which will break the ${\mathfrak {su}}(2)$ R-symmetry down to the Cartan ${\mathfrak u}(1)$. This will become the four dimensional ${\cal N}=1$ $\U(1)$ R-symmetry. Then, we can also turn on a flux supported on the compactification surface for the Cartan generator of the ${\mathfrak {su}}(2)$ global symmetry. Zero value of the flux corresponds to deformations of ${\cal N}=2$ compactifications by giving mass to all the adjoint chiral superfields in ${\cal N}=2$ vector multiplets~\cite{Benini:2009mz}. The ${\cal N}=2$ compactifications correspond to the flux being locked with the Euler characteristic of the surface. Then we have a plethora of other possibilities.

    The simplest such compactification in the $A_1$ case is the one on a two-punctured sphere with one unit of flux, a tube theory.\footnote{ In these notations the ${\cal N}=2$ sphere with three punctures has flux a half.} See {\it e.g.} \cite{Nardoni:2016ffl,Razamat:2019sea,Hwang:2021xyw,Fazzi:2016eec,Agarwal:2015vla}.
    This is given by a bi-fundamental chiral superfield of two puncture $\SU(2)$ symmetries as well as a flip field. The index of this theory in the notations for ${\cal N}=2$ theories we were using till now is,\footnote{Interpreting the ${\cal N}=2$ index as ${\cal N}=1$ one, the ${\cal N}=1$ adjoint chiral superfield in the ${\cal N}=2$ vector multiplet would have R-charge $2$ and hypermultiplets R-charge $0$. Then the bifundamental field of the tube has R-charge $-1$ and the flip field R-charge $+4$.}
    \be 
    {\cal I}_{0,2;{\cal F}=1}(z_1,z_2)=\Gamma_e\left(\frac{t}{\sqrt{p q}} z_1^{\pm1}z_2^{\pm1}\right)\,\Gamma_e\left(\left(\frac{p q}{t}\right)^2\right)\,.
    \ee This index has a simple behavior in the non-relativistic limit. The flip field contribution, $\Gamma_e\left(\left(\frac{p q}{t}\right)^2\right)$, is pole divergent and can be stripped off very similarly to what we do with the Coulomb branch divergences in this limit. However, the bifundamental field leads to finite result,
    \be\label{eq:tubegeneralizedschur}
    \Gamma_e\left(\frac{t}{\sqrt{p q}} z_1^{\pm1}z_2^{\pm1}\right)\;\;\to\;\; \frac{1}{\theta_q(q^{\frac12} z_1^{\pm1}z_2^{\pm1})^\alpha}\,.
    \ee Thus the non-relativistic limit, the generalized Schur limit, of {\it any} ${\cal N}=1$ $A_1$ class ${\cal S}$ compactification built from gluing of ${\cal N}=2$ trinions and the tubes discussed here can be computed after stripping the divergences from Coulomb branch and flip fields and is well defined. The ${\cal N}=1$ class ${\cal S}$ theories which can be obtained this way can be constructed from ${\cal N}=2$ class ${\cal S}$ theories by flipping some of the punctures and then closing them. This gives theories with flux ${\cal F}\geq g-1+\frac{\frak s}2$.  We thus can claim that this 
    limit is well defined beyond ${\cal N}=2$ supersymmetry and in particular the Schur limit can be defined by taking $\alpha=1$, the free particle limit of the Lam\`{e} equation.

    Let us note here that the non-relativistic limit of the two punctured sphere with flux one coincides with the non-relativistic limit of a three punctured sphere with flux half if one switches off the fugacity for the additional puncture. Stating this differently,
    the generalized Schur index of a theory with flux ${\cal F}$ and ${\mathfrak s}$ punctures with all the puncture parameters switched off, depends only on $\frac12\,{\mathfrak s}+{\cal F}$, and not on the two parameter individually.\footnote{This statement generalizes the observation made in \cite{Razamat:2020gcc} about the Schur index of ${\cal N}=1$ theories with flux zero. }

    \ 

    This statement will be generalized in  section \ref{sec:Estring}. The ${\cal N}=2$ supersymmetry is in fact not essential to define the generalized Schur index. One can define it in a variety of ${\cal N}=1$ situations as the non-relativistic limit. One can define a direct analogue of the Schur index as the {\it free (fermionic) particle} limit of the corresponding integrable model.\footnote{For explicit connections of the Schur index of particular theories to free fermions see \cite{Bourdier:2015wda,Bourdier:2015sga}. For closely related relations of free fermions and partition functions of ${\cal N}=2$ SCFTs see \cite{Nekrasov:2003rj}.} 

    \ 

    Before turning our attention to manifestly ${\cal N}=1$ classes of theories let us discuss next a surprising relation of generalized Schur indices of (some of the) theories connected by Coulomb/mass RG flows as manifested in eigenfunctions of the integrable models in the non-relativistic limit.

\section{An example of a higher rank non-relativistic limit}\label{sec:A2generalizedSchur}

Let us discuss an example of a surprising relation of generalized Schur indices of theories related by Coulomb/mass deformations. We consider the rank one Deligne-Cvitanovi\'{c} series of ${\cal N}=2$ SCFTs. See {\it e.g.} \cite{Beem:2019tfp,Banks:1996nj,Dasgupta:1996ij,Sen:1996vd}. This is a sequence of algebras,
\be
{\mathfrak e}_8\;\to\;{\mathfrak e}^*_{7\frac12}\;\to\;{\mathfrak e}_7\;\to\;{\mathfrak e}_6\;\to\;{\mathfrak f}^*_4\;\to\;{\mathfrak d}_4\;\to\;
{\mathfrak g}^*_2\;\to\;{\mathfrak a}_2\;\to\;{\mathfrak a}_1\;\to\;{\mathfrak a}^*_{\frac12}\;\to\;{\mathfrak a}_0\,,
\ee most of which (the ones not marked by $*$) correspond to rank one ${\cal N}=2$ SCFTs.\footnote{See however 
 \cite{Bourget:2020asf} for interesting statements about  ${\mathfrak f}^*_4$ and ${\mathfrak g}^*_2$, and \cite{Cho:2024civ} for interesting statements about ${\mathfrak a}^*_{\frac12}$.} 
This sequence of theories contains the ${\mathfrak d}_4$ SCFT, $\SU(2)$ ${\mathcal N}=2$ SQCD, which has a manifestly ${\cal N}=2$ supersymmetric Lagrangian description. The generalized Schur index of this model, conveniently normalized, is given by \cite{Deb:2025ypl},
\be
	\label{eq:su2alpha}
	{\cal I}_{\mathfrak{d}_4}(q,\alpha)=\frac{(q;q)^{2}}{{\mathtt{N}}(\alpha)}\oint\frac{dz}{4\pi i z} \left(\frac{\Delta(z)(q\, z^{\pm2};q)^2}{(q^{\frac12}z^{\pm1};q)^8}\right)^\alpha\,,\quad 
	\ee where we have defined the normalization,
	\be
	&&\Delta(z)=(1-z^2)(1-z^{-2})\,,\quad{\mathtt{N}}(\alpha)=\oint\frac{dz}{4\pi i z}\Delta(z)^\alpha\,.\;\;\;
\ee
The interesting observation of \cite{Deb:2025ypl} is that 
the other theories in the rank one Deligne-Cvitanovi\'{c} series have generalized Schur indices expressible in terms of the above,
\be\label{eq:delignecvitanovicrelation}
	{\cal I}_{{\mathfrak g}(h_{\mathfrak g}^\vee)}(q,\,\alpha)={\cal I}_{\mathfrak{d}_4}(q,\,\frac{h_{\mathfrak g}^\vee}6\,\alpha)\,.
	\ee Here $h^\vee_{\mathfrak{g}}$ is the dual-Coxeter number for Lie algebra $\mathfrak{g}$. For example, for the ${\mathfrak e}_6$ theory $h^\vee_{\mathfrak{g}}=12$;
    for the ${\mathfrak e}_7$ theory $h^\vee_{\mathfrak{g}}=18$;
    for the ${\mathfrak a}_2$ AD theory $h^\vee_{\mathfrak{g}}=3$; {\it etc}. 

    We will focus here on the ${\mathfrak e}_6$ SCFT and use the above to write its generalized Schur index as ${\cal I}_{\mathfrak{d}_4}(q,\,2\,\alpha)$. On one hand the $\SU(2)$ ${\cal N}=2$ SQCD corresponds to a four punctured sphere and thus we can write this as,
    \be\label{eq:d4side}
    {\cal I}_{\mathfrak{d}_4}(q,\,2\,\alpha)=\sum_\lambda (C^{(A_1)}_{\lambda}(q,2\alpha))^2\psi_\lambda(1;q,2\alpha)^4\,,
    \ee where $\psi_\lambda(z;q,\alpha)$ are eigenfunctions of the $A_1$ elliptic Calogero-Moser model. 

    On the other hand, the ${\mathfrak e}_6$ SCFT can be engineered in class ${\cal S}$ as compactification on a sphere with three maximal punctures of the $A_2$ $(2,0)$ theory. Thus we should be able to write the index as,
    \be\label{eq:e6side}
    {\cal I}_{\mathfrak{d}_4}(q,\,2\,\alpha)=\sum_{\lambda_1\geq \lambda_2} C^{(A_2)}_{\lambda_1,\,\lambda_2}(q,\alpha)\psi_{\lambda_1,\,\lambda_2}(1,1;q,\alpha)^3\,,
    \ee where $\psi_{\lambda_1,\,\lambda_2}(z_1,z_2;q,\alpha)$ are eigenfunctions of the $A_2$ elliptic Calogero-Moser model. We will not review the derivation here of this integrable model explicitly as it is practically the same as for the $A_1$ case. Let us only state that for general $A_{N-1}$ models one of the two basic Hamiltonians becomes a simple shift operator in the non-relativistic limit while the second one becomes a second order elliptic $A_{N-1}$ Calogero-Moser operator,
    \be
   && H \cdot \psi_{\{\lambda_i\}}(\{x_i\})=\\
   &&\;\;\;\;\;\;\left[	-\frac{1}{2}\sum_{i=1}^N \frac{\partial^2}{\partial x_i^2}+g(g-1)\sum_{i<j}\wp(x_i-x_j|\tau)\right]\,\psi_{\{\lambda_i\}}(\{x_i\})=
    {\cal E}_{\{\lambda_i\}}\,\psi_{\{\lambda_i\}}(\{x_i\})\,.\nonumber
    \ee A lot is known about these operators, their kernel functions, and their eigenfunctions. See {\it e.g.} \cite{Hallnas:2024jhj,langmann2010source}.\footnote{Reintroducing factors of mass and $\hbar$ this can be neatly written as, \begin{equation}
		H=-\frac{\hbar^2}{2m}\sum_{i=1}^N \frac{\partial^2}{\partial x_i^2}+\frac{g(g-\hbar)}{m}\sum_{j,k=1,j<k}^M \wp(x_j-x_k|\tau)\,.
	\end{equation}}

   Here we will explicitly find the eigenfunctions and the structure constants to check the equality of the generalized Schur indices expressed  using $A_1$ and $A_2$ eigenfunctions.

   \subsection{Derivation of the structure constants}
  We want to check \eqref{eq:delignecvitanovicrelation} by showing the equality of \eqref{eq:d4side} and \eqref{eq:e6side}.
  We have computed \eqref{eq:d4side} in the previous section so here we will detail how \eqref{eq:e6side} is derived.
  We need eigenfunctions of the $A_2$ elliptic Calogero-Moser model $\psi_{\lambda_1,\,\lambda_2}(z_1,z_2;q,\alpha)$
  and the structure constants $C^{(A_2)}_{\lambda_1,\,\lambda_2}(q,\alpha)$. The former can be computed in a very similar manner to what we did for $A_1$ case and we will detail this in the next sub-section. The structure constants 
  can be derived from the eigenfunctions in a variety of ways. Here we will detail one particular algorithm
  which will exploit the existence of the Argyres-Seiberg duality \cite{Argyres:2007cn}. In fact, this duality was used to derive the first explicit expression for the full index of the ${\mathfrak e}_6$ SCFT \cite{Gadde:2010te}.\footnote{
 An approach to derive eigenfunctions themselves using dualities  was outlined in \cite{Razamat:2013qfa}.}
 We will perform the discussion at the level of the generalized Schur index to avoid cluttered notations, but it holds more generally adjusting appropriately some of the expressions. 

Argyres-Seiberg duality relates the theory obtained by gauging an $\SU(2)$ subgroup of the $\E_6$ global symmetry of the ${\mathfrak e}_6$ SCFT supplemented with additional charged free hypermultiplet matter to ${\cal N}=2$ $\SU(3)$ superconformal QCD. The latter theory having explicit ${\cal N}=2$ Lagrangian, as well as the former gauging with hypermultiplet matter, are easily and directly analyzed at the level of the index. All the information about the index of the ${\mathfrak e}_6$ SCFT can be then extracted from the duality relation between the two descriptions.

    The starting point is the theory corresponding to compactification on a sphere with one minimal and two maximal punctures of the $A_2$ $(2,0)$ theory. This is given by a bi-fundamental hypermultiplet of two maximal puncture $\SU(3)$ symmetries. The index of this theory can be computed, and once we solve for the eigenfunctions of the RS model (or one of its limits) it can be written explicitly as,
	\be
    \label{eq:A2hyptqftexpand}
\prod_{i,j=1}^3\Gamma_e(t^{\frac12}z_iy_j^{-1})=\sum_{\lambda_1=0}^\infty\sum_{\lambda_2=0}^{\lambda_1} \tilde C_{\lambda_1,\lambda_2}\psi_{\lambda_1,\lambda_2}({\bf z})\psi_{\lambda_1,\lambda_2}({\bf y}^{-1})\,,
	\ee with $\tilde C_{\lambda_1,\lambda_2}$ determined by this relation. We do not assume that the functions are  necessarily normalized,
	\be\label{eq:A2normalizedeigenfunctions}
	\oint\left[d{\bf z}\right]_\alpha \psi_{\lambda_1,\lambda_2}({\bf z})\psi_{\lambda'_1,\lambda'_2}({\bf z}^{-1})=n_{\lambda_1,\lambda_2} \delta_{\lambda_1,\lambda'_1}\delta_{\lambda_2,\lambda'_2}\,.
	\ee We have defined,
	\be
	\left[d{\bf z}\right]_\alpha=
	\frac1{3!}\prod_{i=1}^2\frac{dz_i}{2\pi i z_i}\left(\prod_{i\neq j}\theta(z_i/z_j;q)\right)^\alpha\,.
	\ee
    The index of the ${\cal N}=2$ $\SU(3)$ conformal SQCD is obtained by combining two three punctured spheres above: it corresponds to a sphere with two maximal and two minimal punctures \cite{Gaiotto:2009we}.
	Writing the index of the ${\mathfrak e}_6$ SCFT as,
	\be\label{eq:e6index}
    \mathcal{I}_{\mathfrak{e}_6}(q,\alpha)=\sum_{\lambda_1=0}^\infty\sum_{\lambda_2=0}^{\lambda_1} C^{(A_2)}_{\lambda_1,\lambda_2}\psi_{\lambda_1,\lambda_2}({\bf z})\psi_{\lambda_1,\lambda_2}({\bf y})\psi_{\lambda_1,\lambda_2}({\bf x})\,,
	\ee
	and using Argyres-Seiberg duality, we have a relation
	\be
	&&\sum_{\lambda_1=0}^\infty\sum_{\lambda_2=0}^{\lambda_1} C^{(A_2)}_{\lambda_1,\lambda_2}\psi_{\lambda_1,\lambda_2}({\bf x})\psi_{\lambda_1,\lambda_2}({\bf y})\oint\frac{dz}{4\pi i z}\left(\theta(z^{\pm2};q)\right)^\alpha
	\frac{\psi_{\lambda_1,\lambda_2}(1,z,z^{-1})}{(q^{\frac12}z^{\pm1};q)^{2\alpha}}=\\
&&\;\;\;\;\;\;\sum_{\lambda_1=0}^\infty\sum_{\lambda_2=0}^{\lambda_1} (\tilde C_{\lambda_1,\lambda_2})^2n_{\lambda_1,\lambda_2}\psi_{\lambda_1,\lambda_2}({\bf x})\psi_{\lambda_1,\lambda_2}({\bf y})\, .\nonumber
	\ee On the left-hand side we have the $\SU(2)$ gauging with additional matter of the ${\mathfrak e}_6$ SCFT, while the right-hand side represents the index of the SQCD obtained by gluing two free three-punctured spheres. 
    By projecting this relation on $\psi_{\lambda_1,\lambda_2}({\bf y})$ and using orthogonality of the eigenfunctions, we obtain,
	\be
    \label{eq:strucconstA2tqftexpand}
	C^{(A_2)}_{\lambda_1,\lambda_2}=\left[\oint\frac{dz}{4\pi i z}\left(\theta(z^{\pm2};q)\right)^\alpha\frac{\psi_{\lambda_1,\lambda_2}(1,z,z^{-1})}{(q^{\frac12}z;q)^{2\alpha}}\right]^{-1}
	(\tilde C_{\lambda_1,\lambda_2})^2n_{\lambda_1,\lambda_2}\,.
	\ee 
    Thus, knowing the $A_2$ eigenfunctions allows us to derive the structure constants $\tilde C_{\lambda_1,\lambda_2}$ from \eqref{eq:A2hyptqftexpand}, and then determine $C^{(A_2)}_{\lambda_1,\lambda_2}$ from \eqref{eq:strucconstA2tqftexpand}. We can finally verify the relation we are after.
    
    We will derive next the eigenfunctions using instanton calculus again, and then combine everything to check the relation between ${\cal N}=2$ $\SU(2)$ SQCD and the ${\mathfrak e}_6$ SCFT.

        \subsection{$A_2$ Elliptic Jack functions from instanton computations }\label{sec:A2-instanton}

    Let us now explain how the eigenfunctions of the elliptic $A_2$ Calogero-Moser Hamiltonian can be constructed from ramified instanton calculus. The general framework was described in section~\ref{subsec:A1-instanton} for the elliptic $A_{N-1}$ system, and here we specialize it to the case $N=3$. We therefore consider the 4d $\mathcal{N}=4$ $\SU(3)$ gauge theory in the presence of a codimension-two monodromy defect of maximal type with $\rho=[1,1,1]$. The defect is realized by coupling a 2d $T[\SU(3)]$ theory to the 4d bulk.

Following the general prescription \eqref{eq:normalized-defect-partition-function}, the formal wavefunction of the elliptic $A_2$ CM Hamiltonian is
\begin{align}\label{eq:A2-wavefunction}
	\Psi(x_i,a_j,\alpha;q) &= \prod_{i<j}^3\theta(z_i/z_j;q)^\alpha\cdot D(x_i,a_j,\alpha;q) \ , \nonumber \\
	D(x_i,a_j,\alpha;q) &=\prod_{i=1}^3z_i^{a_i}(z_i/z_{i+1})^\alpha \cdot \lim_{\epsilon_2\rightarrow0}\frac{Z^{2d/4d}_{\rm inst}(z_i,a_j,\alpha;q)}{Z^{4d}_{\rm inst}(a_j,\alpha;q)} \ ,
\end{align}
where $a_j$ are the $\SU(3)$ Coulomb branch parameters satisfying $\sum_j a_j=0$, and $\alpha$ is the $\mathcal{N}=2^*$ mass parameter identified with the CM coupling constant. The ramified instanton partition function for the maximal defect,
\begin{align}
	Z^{2d/4d}_{\rm inst}(z_i;q) = Z_{\rho=[1,1,1]}(q_1,q_2,q_3) \ ,
\end{align}
depends on the three fractional instanton fugacities, which are related to the FI parameters of the 2d defect theory as
\begin{align}\label{eq:A2-fugacities}
	q_1 = z_1/z_2 = e^{x_1-x_2}\,,\quad q_2=z_2/z_3 = e^{x_2-x_3}\,,\quad q_3 = (z_3/z_1)\,q = e^{x_3-x_1}q \ .
\end{align}

The partition function admits an expansion in powers of these fractional instanton fugacities. In the NS limit, the normalized partition function takes the form
\begin{align}\label{eq:A2-partition-function}
	&\frac{Z^{2d/4d}_{\rm inst}}{Z^{4d}_{\rm inst}}= \;1 + q_1\frac{\alpha(a_{12}\!+\!\alpha)}{a_{12}\!+\!1}+q_2\frac{\alpha(a_{23}\!+\!\alpha)}{a_{23}\!+\!1}+q_3\frac{\alpha(a_{31}\!+\!\alpha)}{a_{31}\!+\!1} +q_1^2\frac{\alpha(1\!+\!\alpha)(a_{12}\!+\!\alpha)(a_{12}\!+\!\alpha\!+\!1)}{2(a_{12}\!+\!1)(a_{12}\!+\!2)}\nonumber\\
	&+q_2^2\frac{\alpha(1\!+\!\alpha)(a_{23}\!+\!\alpha)(a_{23}\!+\!\alpha\!+\!1)}{2(a_{23}\!+\!1)(a_{23}\!+\!2)}+q_3^2\frac{\alpha(1\!+\!\alpha)(a_{31}\!+\!\alpha)(a_{31}\!+\!\alpha\!+\!1)}{2(a_{31}\!+\!1)(a_{31}\!+\!2)} \nonumber \\
	&+q_1q_2\left(\frac{\alpha^2(a_{12}\!+\!\alpha\!-\!1)(a_{23}\!+\!\alpha)}{a_{12}(a_{23}\!+\!1)} \!-\!\frac{\alpha(a_{12}\!+\!\alpha)(a_{12}\!-\!\alpha\!+\!1)(a_{13}\!+\!\alpha)}{a_{12}(a_{12}\!+\!1)(a_{13}\!+\!1)}\right) + (\text{perms}) + \cdots \ , 
\end{align}
where $a_{ij}\equiv a_i-a_j$, and $({\rm perms})$ denotes the terms obtained by permuting the pairs $(q_i,a_i)$. As in the $A_1$ case, this expression still depends on the continuous Coulomb parameters $a_i$ and does not define an $L^2$-normalizable eigenfunction until a quantization condition is imposed.

To obtain the physical eigenfunctions, we impose the quantization condition in \cite{Nekrasov:2009rc} that reads
\begin{align}\label{eq:A2-quantization}
	a_1-a_2 = -\alpha - n_2\,,\qquad a_2-a_3 = -\alpha-n_1\,,\qquad n_1,n_2\in\mathbb{Z}_{\geq0}\,.
\end{align}
The pair $(n_1,n_2)$ labels the discrete spectrum of the elliptic $A_2$ Calogero-Moser Hamiltonian through the highest weight $(\lambda_1,\lambda_2)=(n_1+n_2,n_2)$ of the corresponding $\SU(3)$ representation. These provide the properly quantized eigenfunctions of the elliptic $A_2$ CM Hamiltonian.
In the trigonometric limit $q\to 0$, the resulting eigenfunctions reduce to the $A_2$ Jack polynomials, while for finite $q$ they are deformed to elliptic Jack functions which is a $q$-series deformation of the Jack polynomials.

Expanding the quantized eigenfunctions in powers of $q$ and expressing them in terms of $\SU(3)$ characters, we obtain the first few eigenfunctions $D_{\lambda_1,\lambda_2}$ as
\begin{align}\label{eq:A2-eigenfunctions}
    D_{\lambda_1,\lambda_2} \equiv& D(x_i,a_j,\alpha;q)\big|_{a_1-a_2\rightarrow-\alpha-\lambda_2,a_2-a_3\rightarrow -\alpha-\lambda_1+\lambda_2} \ , \nonumber \\ 
	D_{0,0} =& 1+q\tfrac{3\alpha\left(\alpha(\alpha+1)\chi_{\bf 8}+2\alpha^2+4\alpha-2\right)}{(2\alpha+1)(\alpha+1)}+q^2\bigg[\tfrac{3\alpha^2(1\!+\!3\alpha)((1\!+\!\alpha)(\chi_{\bf 27}+\chi_{\bf 3}+\chi_{\bf \bar{3}})\!-\!2(\chi_{\bf 3}+\chi_{\bf \bar{3}})}{4(1+\alpha)(1+2\alpha)}   \\
	&\quad \ \ +  \tfrac{6\alpha^2(1+3\alpha^2+23\alpha^3+9\alpha^4)\chi_{\bf 8}}{(1+\alpha)(1+2\alpha)^3}-\tfrac{3\alpha(24-26\alpha+27\alpha^2+623\alpha^3-271\alpha^4-1113\alpha^5-500\alpha^6-60\alpha^7)}{4(1+\alpha)^2(2+\alpha)(1+2\alpha)^3}\bigg]+\mathcal{O}(q^3), \nonumber \\
D_{1,0} =& \chi_{\bf 3}\!+\!q\!\bigg[\!\tfrac{\alpha(1+3\alpha)\,\chi_{\bf 15}}{2(1+\alpha)}\!+\!\tfrac{\alpha(3\alpha^2+6\alpha-1)\,\chi_{\bf \bar{6}}}{2(1+\alpha)^2}\!+\!\tfrac{(18-39\alpha-\alpha^2+37\alpha^3+9\alpha^4)\,\chi_{\bf 3}}{2(1+\alpha)^2(2+\alpha)}\!\bigg] \!\!+\! q^2\!\bigg[\!\tfrac{\alpha(1 + 3\alpha)(-1 +\alpha +7\alpha^2+ 3\alpha^3)\chi_{\bf 24}}{2(1 + \alpha)^2(3 + 2\alpha)}\nonumber \\
&\qquad+\tfrac{\alpha(1 + 3\alpha)(2 + 3\alpha)\left((1+\alpha)\chi_{\bf 42}-(1 - \alpha)\chi_{\bf 15'})\right)}{4(1+\alpha)(3 + 2\alpha)}+\tfrac{\alpha(18 + 61\alpha - 31\alpha^2 + 19\alpha^3 + 269\alpha^4 + 204\alpha^5 + 36\alpha^6)\chi_{\bf 15}}{4(1 + \alpha)^4(2 + \alpha)} \nonumber \\
&\qquad+ \tfrac{(648 - 2886\alpha - 807\alpha^2 + 14010\alpha^3 + 5186\alpha^4 - 12352\alpha^5 - 3300\alpha^6 + 7978\alpha^7 + 5518\alpha^8 + 1266\alpha^9 + 99\alpha^{10})\chi_{\bf 3}}{4(1 + \alpha)^5(2 + \alpha)(3 + \alpha)(3 + 2\alpha)}  \nonumber \\
&\qquad+\tfrac{\alpha(-50 + 503\alpha + 711\alpha^2 - 654\alpha^3 - 337\alpha^4 + 1507\alpha^5 + 1557\alpha^6 + 540\alpha^7 + 63\alpha^8)\chi_{\bf \bar{6}}}{4(1 + \alpha)^5(2 + \alpha)(3 + 2\alpha)}\bigg]+\mathcal{O}(q^3) \ , \nonumber 
\end{align}
\begin{align}
    D_{2,0} =& \chi_{\bf 6}\!+\!\tfrac{\alpha-1}{\alpha+1}\chi_{\bf \bar{3}}\!+\!q\bigg[\tfrac{\alpha(2+3\alpha)\,\chi_{\bf 24}}{3+2\alpha}\!-\!\tfrac{2\alpha(1-\alpha-7\alpha^2-3\alpha^3)\,\chi_{\bf \bar{15}}}{(1+\alpha)^2(3+2\alpha)}\!+\!\tfrac{4\alpha(-3 - 15\alpha - 20\alpha^2 + 29\alpha^3 + 79\alpha^4 + 44\alpha^5 + 6\alpha^6)\chi_{\bf 6}}{(1 + \alpha)^3(3 + \alpha)(1 + 2\alpha)(3 + 2\alpha)} \nonumber \\
	&\qquad+\tfrac{4\alpha(12 + 45\alpha + 25\alpha^2 - 39\alpha^3 + 31\alpha^4 + 40\alpha^5 + 6\alpha^6)\chi_{\bf \bar{3}}}{(1 + \alpha)^3(3 + \alpha)(1 + 2\alpha)(3 + 2\alpha)}\bigg]+ \mathcal{O}(q^2) \ , \nonumber \\
	D_{2,1} =& \chi_{\bf 8}+\tfrac{2(\alpha-1)}{2\alpha+1}\!+\!q\bigg[\tfrac{\alpha(2+3\alpha)\,\chi_{\bf 27}}{3+2\alpha}\!+\!\tfrac{\alpha(2+10\alpha+3\alpha^2)\big(\chi_{\bf 10}+\chi_{\bf \bar{10}}\big)}{(2+\alpha)(3+2\alpha)}\!-\!\tfrac{2(1 - \alpha)(-12 - 46\alpha + 17\alpha^2 + 155\alpha^3 + 30\alpha^4)\chi_{\bf 8}}{(2 + \alpha)(1 + 2\alpha)^2(3 + 2\alpha)} \nonumber \\
	&\qquad +\tfrac{-48 + 64\alpha + 432\alpha^2 - 477\alpha^3 + 128\alpha^4 + 36\alpha^5}{(2 + \alpha)(1 + 2\alpha)^2(3 + 2\alpha)}\bigg]+\mathcal{O}(q^2) \ , \nonumber \\
	D_{1,1} =& D_{1,0}({\bf \bar{z}}) \ , \quad D_{2,2} = D_{2,0}({\bf \bar{z}}) \ ,\nonumber
\end{align}
where $\chi_{\bf r}$ denotes the $\SU(3)$ character in the representation ${\bf r}=(\lambda_1,\lambda_2)$ with fugacities ${\bf z}=\{z_1,z_2\}$. In particular, $D_{n_1+n_2,n_2}$ and $D_{n_1+n_2,n_1}$ are related by charge conjugation ${\bf z}\to\bar{\bf z}$, as expected from the symmetry of the Hamiltonian.  One can verify that  eigenfunctions $D$ agree (under the relation \eqref{eq:A2-wavefunction} and up to $(q,\alpha)$-dependent proportionality constant) with wavefunctions $\psi$ obtained by directly solving the elliptic $A_2$ CM differential equation with the explicit results collected in appendix \ref{app:A2technicaldetails}.

Finally, the eigenvalues are determined by the chiral ring vev of the $\SU(3)$ theory. The general relation \eqref{eq:chiral-ring} for $N=3$ gives
\begin{align}\label{eq:A2-eigenvalue}
	\mathcal{E}_{\SU(3)} = \langle\mathrm{Tr}\,a^2_{\SU(3)}\rangle - 3\alpha(\alpha-1)\frac{E_2(q)}{24}\,,
\end{align}
where $\langle\mathrm{Tr}\,a^2_{\SU(3)}\rangle$ is the expectation value of the second Casimir chiral ring operator. For general Coulomb branch parameters, the corresponding formal eigenvalue is
\begin{align}
	&\mathcal{E}_{\SU(3)}+6\alpha(\alpha-1)\frac{E_2(q)}{24}-\frac{\alpha(\alpha-1)}{8} \\
	&= \;\sum_i^3\frac{a_i^2}{2}- q\!\left[\frac{(\alpha\!-\!1)\alpha(a_{12}\!+\!\alpha)(a_{13}\!+\!\alpha)(a_{12}\!-\!\alpha\!+\!1)(a_{13}\!-\!\alpha\!+\!1)}{a_{12}a_{13}(a_{12}+1)(a_{13}+1)} \!+\!({\rm perm})\right]+ \mathcal{O}(q^2)\,. \nonumber 
\end{align}
Imposing the quantization condition \eqref{eq:A2-quantization} in this expression then yields the discrete spectrum of the elliptic $A_2$ CM model, which can be checked against the eigenvalues obtained from the differential equation.

\ 

Once we have the eigenfunctions we can use the algorithm developed in the previous subsection to derive the structure constants and then compute the generalized index of the $\mathfrak{e}_6$ SCFT. Since the technical details are cumbersome and lengthy, we defer the details to appendix \ref{app:A2technicaldetails}. Here let us quote the result of the index \eqref{eq:e6index} after setting all the flavor fugacities to one,
\begin{equation}\label{eq:e6generalzedSchurIndex}
	\mathcal{I}_{\mathfrak{e}_6}(q,\alpha)=1+\frac{2  (10 \alpha -1) (1+12 \alpha )}{1+2\alpha }q+\frac{-1-3 \alpha +58 \alpha ^2+600 \alpha ^3+14400 \alpha ^4}{(1+2\alpha ) (1+\alpha )}q^2+{\cal O}(q^3)~.
\end{equation}
This is equal to the index of $\SU(2)$ conformal SQCD given in \eqref{eq:d4result} with the rescaling of $\alpha$ by $2$ as expected from \eqref{eq:delignecvitanovicrelation}.

    \section{E-string compactifications and the Inozemtsev model}\label{sec:Estring}

      Since the non-relativistic model is related to the generalized Schur index for $(2,0)$ ${\cal N}=2$ compactifications, it is natural to ask whether similar non-relativistic limits play an interesting role in other compactifications from six dimensions down to four. Maybe the simplest such compactification is that of the E-string $(1,0)$ SCFT. 
    Like the $A_1$ $(2,0)$, the compactifications of rank one E-string have been thoroughly understood in recent years: the theories one obtains in four dimensions for arbitrary surfaces and arbitrary fluxes are vanilla Lagrangian models \cite{Kim:2017toz,Razamat:2020bix,Kim:2023qbx,Razamat:2022gpm,Razamat:2019ukg}. On the other hand it is also well understood \cite{Nazzal:2018brc,Nazzal:2021tiu,Chen:2021ivd} that the corresponding integrable model is the $BC_1$ van~Diejen model \cite{vanDiejen1994Integrability}. The eight parameters of the van~Diejen model are mapped to the eight fugacities of the Cartan subgroup of the $\E_8$ symmetry group of the rank one E-string theory. Some of these statements have generalizations for higher rank E-string theory. For example, many compactifications on low genus surfaces with two or less punctures are understood \cite{Pasquetti:2019hxf,Hwang:2021xyw}. The integrable model corresponding to the rank-$Q$ E-string is expected to be the $BC_Q$ van~Diejen model with the nine parameters mapping to the Cartan subgroup fugacities of the $\SU(2)\times \E_8$ symmetry of the model with $Q>1$.  The van~Diejen model has an interesting non-relativistic limit, the Inozemtsev model. We will explore here the physics of this limit in our index/compactification framework.\footnote{Let us mention here that the van~Diejen model has also a limit more akin to the Macdonald limit of the RS model. In this limit the eigenfunctions are given by Koornwinder polynomials \cite{MR1199128}. This limit was commented upon in the context of E-string compactifications in \cite{Nazzal:2018brc} and also appeared in twisted class ${\cal S}$ compactifications in \cite{Mekareeya:2012tn}. See also \cite{Ren:2025tvx}. The Inozemtsev model has been discussed recently through its connection to the Seiberg-Witten curve of $\SU(2)$ $N_f=4$ SQCD \cite{Argyres:2021iws}.} In this section instead of focusing on detailed discussion of limits of eigenfunctions, which are much more involved here than in the  RS case, we will only discuss non-relativistic limits of physical theories. 

    \subsection{Rank one case}

        \

 \begin{figure}
 \center
 \includegraphics[width=0.4\textwidth]{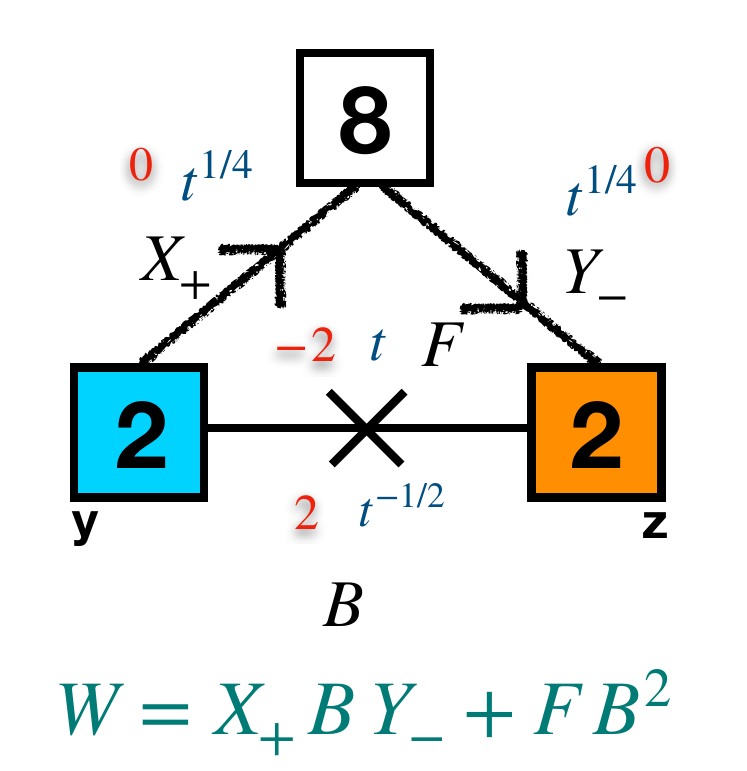} 
 \caption{The tube theory for simplest E-string compactification. The numbers in red denote a choice of R-symmetry assignment consistent with the superpotential. The theory has a $\U(1)_t$ symmetry with charges of various fields encoded in the powers of fugacity $t$. }
 \label{F:TubeE}
 \end{figure}

    Let us start by analyzing the simplest model obtained by compactifying the rank one E-string to four dimensions: compactification on a two-punctured sphere with minimal flux preserving $\E_7\times \U(1)$ symmetry given by a WZ model \cite{Kim:2017toz} of Figure \ref{F:TubeE}.

    The basic technical idea of taking the non-relativistic limit in this system is as follows. First, we have the elliptic Gamma function identity,
    \be\label{eq:SplitToEight}
    \Gamma_e(z^2)=\Gamma_e(\pm z)\,\Gamma_e(\pm \sqrt{q} z)\,\Gamma_e(\pm \sqrt{p} z)\,\Gamma_e(\pm \sqrt{p q} z)\,.
    \ee The elliptic Gamma function also has the following periodicity property,
    \be
    \Gamma_e( p z)=\theta_q(z)\, \Gamma_e(z)\,.
    \ee Using this property one can obtain that,
    \be\label{eq:RatioOfqLimit}
    \lim_{p\to 1} \frac{\Gamma_e(p^\alpha z)}{\Gamma_e(z)}=\theta_q(z)^\alpha\,.
    \ee For integer $\alpha$ this statement is immediate and for general $\alpha$ it can be carefully derived by considering the log of the equality and staying away from the singularities of the ratio \cite{Rains2009Limits}.
    This is also the basic property which leads to the generalized Schur index of \cite{Deb:2025ypl}.
    
    Let us apply these properties for the index of the E-string compactifications.  We refer to Figure \ref{F:TubeE} for definitions of the fields.
    The fields $X_+$ have weight $t^{\frac14}a_i$ ($\prod_{i=1}^8 a_i=1$) in the index and are in a doublet of the $\SU(2)$ puncture symmetry. 
    We take the following limit,
    \be
    t^{\frac14}a_{1,2}=q^{\frac12}\, e^{\epsilon\, v_{1,2}}\,,\qquad
     t^{\frac14}a_{3,4}= e^{\epsilon\, v_{3,4}}\,,\qquad
      t^{\frac14}a_{5,6}=-q^{\frac12}\, e^{\epsilon\, v_{5,6}}\,,\qquad
       t^{\frac14}a_{7,8}=-\, e^{\epsilon\, v_{7,8}}\,,\qquad
    \ee and $p=e^{\epsilon}$ with $\epsilon\to 0$. The parameters $v_i$, $q$, and $t$ are kept finite and are constrained to be,
    \be
    t=q\,e^{\frac\epsilon2\sum_{i=1}^8 v_i}\,,\qquad \prod_{i=1}^8 a_i=1\,.
    \ee
    We see the motivation for this parametrization is the analogy with the generalized Schur index of $(2,0)$ compactifications. Note that the weight of the fields $Y_-$ is $t^{\frac14}a_i^{-1}$. Thus defining $v=\frac14\sum_{i=1}^8 v_i$,
  \be\label{eq:LimitOfParameters}
    t^{\frac14}a_{1,2}^{-1}=e^{\epsilon\,(v- v_{1,2})}\,,\quad
     t^{\frac14}a_{3,4}^{-1}= q^{\frac12}e^{\epsilon\,(v- v_{3,4})}\,,\quad
      t^{\frac14}a_{5,6}^{-1}=- e^{\epsilon\,(v- v_{5,6})}\,,\quad
       t^{\frac14}a_{7,8}^{-1}=-q^{\frac12}\, e^{\epsilon\, (v-v_{7,8})}\,,\nonumber\\
    \ee 
    To obtain a finite limit we actually have to consider both the contribution of say $X_+$ and the gauge fields that one would introduce while gauging the puncture symmetry, $1/\Gamma_e(z^{\pm2})$. Taking the limit for the contributions of the two octet fields we obtain,
    \be 
    &X_+\,:\;& \frac{\prod_{i=1}^8\Gamma_e(t^{\frac14}a_iz^{\pm1})}{\Gamma_e(z^{\pm2})}\;\;\to\;\; \frac{\theta_q(z^{\pm1})^{v_3+v_4}\theta_q(-z^{\pm1})^{v_7+v_8}\theta_q(q^{\frac12}z^{\pm1})^{v_1+v_2}\theta_q(-q^{\frac12}z^{\pm1})^{v_5+v_6}}{\theta_q(z^{\pm2})^{\frac12}}\,.\nonumber\\
\ee To see this we use \eqref{eq:SplitToEight} and write,
\be
\frac{\prod_{i=1}^8\Gamma_e(t^{\frac14}a_iz^{\pm1})}{\Gamma_e(z^{\pm2})}=\frac{\prod_{i=1}^8\Gamma_e(t^{\frac14}a_iz^{\pm1})}{\Gamma_e(\pm z^{\pm1})\Gamma_e(\pm \sqrt{q}z^{\pm1})\Gamma_e(\pm\sqrt{p} z^{\pm1}) \Gamma_e(\pm\sqrt{p q} z^{\pm1})}\,,
\ee obtaining a product of ratios of the form appearing in \eqref{eq:RatioOfqLimit}. Applying the identity \eqref{eq:RatioOfqLimit} to pairs of elliptic Gamma functions in the numerator and denominator we obtain the quoted result. 

For the other octet field we get a similar result but with different powers, 
\be 
     &Y_-\,:\;& \frac{\prod_{i=1}^8\Gamma_e(t^{\frac14}a_i^{-1}z^{\pm1})}{\Gamma_e(z^{\pm2})}\;\;\to\;\; \\
     && \frac{1}{\theta_q(z^{\pm2})^{\frac12}}\frac{\left(\theta_q(z^{\pm1})\theta_q(-z^{\pm1})\theta_q(q^{\frac12}z^{\pm1})\theta_q(-q^{\frac12}z^{\pm1})\right)^{2v}}{\theta_q(z^{\pm1})^{v_7+v_8}\theta_q(-z^{\pm1})^{v_3+v_4}\theta_q(q^{\frac12}z^{\pm1})^{v_5+v_6}\theta_q(-q^{\frac12}z^{\pm1})^{v_1+v_2}}\,.\nonumber
    \ee The two octets can be thought of as a ${\cal N}=1$ E-string version of ${\cal N}=2$ moment map operators associated to the two different punctures of the tube theory. As the two punctures are different, one obtains octets with slightly different symmetry assignments leading to the different powers in the limit.
    The contributions of the  remaining  bifundamental field $B$ is,
    \be 
    &B\,:\;&\Gamma_e(q p t^{-\frac12}y^{\pm1}z^{\pm1})\;\;\to\;\; \frac1{\theta_q(q^{\frac12}y^{\pm1}z)^{2v-1}}\,.
    \ee Finally combining the flip field $F$ together with the Cartan generator of the $\SU(2)$ gauge field associated to the puncture, we have
    \be
    (q;q)(p;p)\Gamma_e(\frac{t}{pq})\,,
    \ee which leads to a pole divergency for each gauge group and can be conveniently stripped off. 
    We can combine the tube theories of Figure \ref{F:TubeE} into tubes with higher values of flux and to tori with flux by integrating over the $\SU(2)$ parameters associated to the punctures \cite{Kim:2017toz}. 

    We conclude in particular that the Inozemtsev limit is completely well defined for tori theories preserving $\E_7\times \U(1)$ flux. For two punctures spheres with such a flux we need to normalize the index by the contributions of background vector fields for the limit to be well defined. Although the limit was defined using eight parameters $v_i$, only four independent combinations appear in the final result,
    \be
    v_1+v_2=s_1\,,\qquad v_3+v_4=s_2\,,\qquad v_5+v_6=s_3\,,\qquad v_7+v_8=s_4\,,\qquad v=\frac14\sum_{i=1}^4s_i\,.\;\;\;\;\;
    \ee
    Notice that technically in order for the limit to exist it was important that for every gauging we have an octet of fundamental fields. This technical fact physically means that the limit is easily defined for tori/tube theories with  $\E_7\times \U(1)$ flux. For other values of flux not every gauging comes with an octet of fundamental fields \cite{Kim:2017toz}. However, as for the class ${\cal S}$ theories the non-relativistic index was  well defined for a range of values of flux ({\it i.e.} also for ${\cal N}=1$ compactifications), we expect that also here one should be able to make sense of the limit for wider options of flux and also for higher genus compactifications. We will not discuss it here further but rather focus on understanding the limit for general tori and tubes with the $\E_7\times \U(1)$ flux. 
    We leave studying the Inozemtsev limit for more general compactifications for future work.

\

Let us consider limits of indices of several gauge theories corresponding to tubes with higher flux and tori. First, we combine two basic tubes to a tube of flux one by gauging a diagonal $\SU(2)$ symmetry. As a gauge theory this is just an $\SU(2)$ model with $N_f=6$. The Inozemtsev limit of the index is then,
\be\label{eq:TubeFluxOne}
I(s_1,s_2,s_3,s_4)=\oint\frac{dz}{2\pi i z} \frac{\theta_q(z^{\pm1})^{s_2}\theta_q(-z^{\pm1})^{s_4}\theta_q(q^{\frac12}z^{\pm1})^{s_1}\theta_q(-q^{\frac12}z^{\pm1})^{s_3}}{\theta_q(q^{\frac12}y_1^{\pm1}z)^{2v-1}\theta_q(q^{\frac12}y_2^{\pm1}z)^{2v-1}\theta_q(z^{\pm2})^{\frac12}}\,.
\ee Taking a special further limit of $s_i=v$ we obtain that,
\be 
I(v,v,v,v)=\oint\frac{dz}{2\pi i z} \left(\frac{\theta_q(z^{\pm2})}{\theta_q(q^{\frac12}y_1^{\pm1}z)^{2}\theta_q(q^{\frac12}y_2^{\pm1}z)^{2}}\right)^{v-\frac12}\,.
\ee Identifying $v-\frac12\equiv \alpha$ this is exactly proportional to the generalized Schur index of $\SU(2)$ ${\cal N}=2$ SQCD with $N_f=4$. Here, we have used the fact that,
\be
\theta_q(z)\, \theta_q(-z)\, \theta_q(q^{\frac12}z)\, \theta_q(-q^{\frac12}z)=\theta_q(z^2)\,.
\ee Mathematically, we can view thus the expression \eqref{eq:TubeFluxOne} as a refinement of the Schur limit in ${\cal N}=2$ where the vector multiplet contribution is split. The $N_f=6$ $SU(2)$ SQCD and ${\cal N}=2$ $SU(2)$ SQCD are physically different theories and thus the relation between the limits of the indices of the two is not an indication of a duality. We comment on possible reasons for this equality (and similar ones appearing below) in section \ref{sec:summaryanddiscussion}.

As another example, consider gluing the basic tube to itself. We will obtain torus of flux a half. Since we glue two different punctures together some of the symmetry is broken  \cite{Kim:2017toz}. This implies in the limit that we have to identify $s_4=s_2$ and $s_3=s_1$.
In fact the theory one obtains is $\SU(2)$ $N_f=4$ ${\cal N}=2$ SQCD, albeit we are taking a different limit than the generalized Schur index.
We obtain the following result (dropping over-all factors),
\be\label{eq:TorusFluxHalf}
I(s_1,s_2)&=&\oint\frac{dz}{2\pi i z} \frac{\theta_q(z^{\pm1})^{s_2}\theta_q(-z^{\pm1})^{s_2}\theta_q(q^{\frac12}z^{\pm1})^{s_1}\theta_q(-q^{\frac12}z^{\pm1})^{s_1}}{\theta_q(q^{\frac12}z^{2})^{2v-1}\theta_q(z^{\pm2})^{\frac12}}=\\
&&\qquad \qquad \oint\frac{dz}{2\pi i z} \frac{\theta_{q^2}(z^{\pm2})^{s_2}\theta_{q^2}(q z^{\pm2})^{s_1}}{\theta_q(q^{\frac12}z^{2})^{2v-1}\theta_q(z^{\pm2})^{\frac12}}\,.\nonumber
\ee
Note that $\theta_q(z)\theta_q(-z)=\theta_{q^2}(z^2)$. Again, setting further $s_1=s_2$ we obtain,
\be
I(v,v)=\oint\frac{dz}{2\pi i z} \left(\frac{\theta_{q}(z^{\pm2})}{\theta_q(q^{\frac12}z^{\pm2})}\right)^\alpha\,.
\ee This is in fact the generalized Schur limit of the ${\cal N}=4$ $\SU(2)$ SYM. Again \eqref{eq:TorusFluxHalf} can be viewed as a refinement of this Schur index. We stress that although the physical theory here is $\SU(2)$ $N_f=4$ ${\cal N}=2$ SQCD, the Inozemtsev limit of its index with specialized parameters is the same as the generalized Schur partition function for the ${\cal N}=4$ $\SU(2)$ SYM.

\

These two examples can be immediately generalized. Taking $s_i=v$ E-string on torus with flux $n$ gives the same expression as the generalized index of $A_1$ class ${\cal S}$ theory on a torus with $2n$ minimal punctures. Similarly  E-string on tube with flux $n$ has  the same Inozemtsev index as the generalized Schur index of $A_1$ class ${\cal S}$ theory on a sphere with two maximal punctures and with $2n$ minimal punctures. In particular, taking $\alpha=1$ the index of a tube theory has natural expansion in terms of Schur polynomials. We thus observe that different limits of indices of different theories sometimes coincide here similarly to what we observed in section \ref{sec:N1Schur}.

    \subsection{Higher rank cases}

    The discussion here has a rather straightforward generalization to rank $Q$ E-string theory. The main point is that when two theories are glued together along a maximal puncture, for general $Q$ there are three types of fields one needs to consider  \cite{Pasquetti:2019hxf}. First, the vector multiplet and an octet of fundamental fields of $\USp(2Q)$, which are a direct generalization of what we had for rank one ($Q=1$). However, in addition we also have a field in traceless antisymmetric irrep of $\USp(2Q)$ (which does not exist for $Q=1$). The weight of this field is $p q/h$ in our conventions: it has R-charge two and is charged under new $\U(1)_h$ symmetry. To take the non-relativistic limit we parametrize the various fugacities following \eqref{eq:LimitOfParameters} as before and in addition take $h= e^{\epsilon\,\lambda}$. We take then the limit of $\epsilon\to 0$. The contributions of the three types of fields give us,   
\be 
&&\frac{\prod_{j=1}^Q\prod_{i=1}^8\Gamma_e(t^{\frac14}a_iz_j^{\pm1})\prod_{1\leq l<m\leq Q}\Gamma_e(h z_l^{\pm1}z_m^{\pm1})}{\prod_{j=1}^Q\Gamma_e(z_j^{\pm2})\prod_{1\leq l<m\leq Q}\Gamma_e(z_l^{\pm1}z_m^{\pm1})}\;\;\to\;\; \\
&&\;\;\;\prod_{j=1}^Q\frac{\theta_q(z_j^{\pm1})^{v_3+v_4}\theta_q(-z_j^{\pm1})^{v_7+v_8}\theta_q(q^{\frac12}z_j^{\pm1})^{v_1+v_2}\theta_q(-q^{\frac12}z_j^{\pm1})^{v_5+v_6}}{\theta_q(z_j^{\pm2})^{\frac12}}\prod_{1\leq l<m\leq Q}
\theta_q(z_l^{\pm1}z_m^{\pm1})^\lambda\,.\nonumber
\ee 
In particular we will have five independent parameters, four $s_i$ as for rank $Q=1$, and the additional parameter $\lambda$.

\ 

Let us consider next indices of some concrete  theories obtained in rank $Q$ E-string compactifications and discuss the limit for them. Some compactifications were studied in \cite{Pasquetti:2019hxf,Hwang:2021xyw}. However,
for higher rank these are more complicated to analyze as even the simplest compactifications involve non-trivial gauge interactions. The complexity is similar to the one encountered analyzing rank one E-string compactications with general flux or higher genus. 
On the other hand a certain relevant deformation of compactifications on tubes and tori which was studied in \cite{Kim:2017toz} is a straightforward generalization of our rank one discussion. This deformation was also discussed in \cite{Pasquetti:2019hxf} following the work of \cite{Rains2018InterpolationKernel}. The relevant deformation at hand breaks the $\U(1)_h$ symmetry and thus we do not have the freedom of the additional parameter. In particular,
when we glue two punctures after the deformation the weight of the antisymmetric field is fixed to be the same as that of the flip field $F$, $t/ ( p q)$. Another way to phrase this is that the weight $h$ is constrained to satisfy $h=t/(p q)$. In particular in the limit we obtain,
\be 
&&\frac{\prod_{j=1}^Q\prod_{i=1}^8\Gamma_e(t^{\frac14}a_iz_j^{\pm1})\prod_{1\leq l<m\leq Q}\Gamma_e(\frac{t}{p q}z_l^{\pm1}z_m^{\pm1})}{\prod_{j=1}^Q\Gamma_e(z_j^{\pm2})\prod_{1\leq l<m\leq Q}\Gamma_e(z_l^{\pm1}z_m^{\pm1})}\;\;\to\;\; \\
&&\;\;\;\prod_{j=1}^Q\frac{\theta_q(z_j^{\pm1})^{v_3+v_4}\theta_q(-z_j^{\pm1})^{v_7+v_8}\theta_q(q^{\frac12}z_j^{\pm1})^{v_1+v_2}\theta_q(-q^{\frac12}z_j^{\pm1})^{v_5+v_6}}{\theta_q(z_j^{\pm2})^{\frac12}}\prod_{1\leq l<m\leq Q}
\theta_q(z_l^{\pm1}z_m^{\pm1})^{2v-1}\,.\nonumber
\ee That is $\lambda$ is locked to be $2v-1$. With this deformation the basic tube is again a very simple WZ model: it is just a bifundamental of two $\USp(2Q)$ symmetries associated to the punctures with the rest given by the fields we have already discussed. 

\ 

As two examples we will again consider the simplest interacting tube and the torus with minimal flux. The index of the former is proportional to,
\be 
\oint\left[\prod_{j=1}^Q\frac{dz_j}{2\pi i z_j} \frac{\theta_q(z_j^{\pm1})^{s_2}\theta_q(-z_j^{\pm1})^{s_4}\theta_q(q^{\frac12}z_j^{\pm1})^{s_1}\theta_q(-q^{\frac12}z_j^{\pm1})^{s_3}}{\theta_q(q^{\frac12}z_j)^{4Q(2v-1)}\theta_q(z_j^{\pm2})^{\frac12}}\right]\prod_{1\leq l<m\leq Q}
\theta_q(z_l^{\pm1}z_m^{\pm1})^{2v-1}\,.\;\;\;\;\;\;
\ee Here for simplicity we set fugacities of the two $\USp(2Q)$ symmetries to one. Choosing as before all $s_i$ to be the same we obtain,
\be 
\oint\prod_{j=1}^Q\frac{dz_j}{2\pi i z_j} \left[\prod_{j=1}^Q\frac{\theta_q(z_j^{\pm2})}{\theta_q(q^{\frac12}z_j)^{8Q}}\prod_{1\leq l<m\leq Q}
\theta_q(z_l^{\pm1}z_m^{\pm1})^2\right]^{v-\frac12}\,.
\ee 

\ 

For the torus with minimal flux as for the rank one case only two parameters remain and we can write,
\be 
\oint\prod_{j=1}^Q\frac{dz_j}{2\pi i z_j} \frac{\theta_{q^2}(z_j^{\pm2})^{s_2}\theta_{q^2}(qz_j^{\pm2})^{s_1}}{\theta_q(q^{\frac12}z_j^{2})^{2v-1}\theta_q(z_j^{\pm2})^{\frac12}}\prod_{1\leq l<m\leq Q}
\frac{\theta_q(z_l^{\pm1}z_m^{\pm1})^{2v-1}}{\theta_q(q^{\frac12}z_l^{\pm1}z_m^{\pm1})^{2v-1}}\,.\nonumber
\ee Taking further $s_1=s_2$ we obtain,
\be 
\oint\prod_{j=1}^Q\frac{dz_j}{2\pi i z_j} \left(\prod_{j=1}^Q\frac{\theta_{q}(z_j^{\pm2})}{\theta_q(q^{\frac12}z_j^{\pm2})}\prod_{1\leq l<m\leq Q}
\frac{\theta_q(z_l^{\pm1}z_m^{\pm1})^2}{\theta_q(q^{\frac12}z_l^{\pm1}z_m^{\pm1})^2}\right)^\alpha\,.
\ee 

\

Finally, let us comment on the relevant integrable model. 
The integrable model in terms of eigenfunctions of which the indices of the E-string compactifications are expected to be naturally written is the $BC_Q$ van~Diejen model, as we mentioned before. The van~Diejen model depends on nine parameters, the fugacities for the Cartan generators of the $\E_8\times \SU(2)$ global symmetry, the $a_i$, $t$, and $h$ in our discussion. Inozemtsev model is obtained by starting with the van~Diejen model and scaling parameters as in \eqref{eq:LimitOfParameters} and $h=e^{\epsilon\lambda}$, leaving us with five parameters. The model has the following form,
\be\label{eq:InozemtsevModel} 
 && H_{\text{Inoz.}} \cdot \psi_{\{\lambda_i\}}(\{x_i\})=\left[	-\frac{1}{2}\sum_{i=1}^Q \frac{\partial^2}{\partial x_i^2}+\lambda(\lambda-1)\sum_{i<j}\left(\wp(x_i-x_j|\tau)+\wp(x_i+x_j|\tau)\right)+\right.\nonumber\\
&&\;\;\;\;\;\;\left.\frac12\sum_{\ell=0}^3g_\ell(g_\ell-1)\sum_{i=1}^Q\wp(x_i+\omega_\ell|\tau)\right]\,\psi_{\{\lambda_i\}}(\{x_i\})={\cal E}_{\{\lambda_i\}}\,\psi_{\{\lambda_i\}}(\{x_i\})
    \,.
\ee Here $\omega_0=0$, $\omega_1=\frac12$, $\omega_2=\frac\tau2$, and $\omega_3=\frac{1+\tau}2$ are half periods of the elliptic curve defined by $q$.\footnote{Note that for $Q=1$ and $g_{i}=g$ the model is equivalent to the Lam\'{e} equation. To show this one first notices that $\sum_{\ell=0}^3\wp(x+\omega_\ell|\tau)=4\wp(2x|\tau)$ and then rescales $2x\to x$. We have also observed that when all couplings are equal, the vector multiplet measure coincides with that of the $A_1$ $(2,0)$ case. Similarly for $Q>1$ setting all $g_i=0$ while keeping $\lambda$ one obtains the $D_Q$ elliptic Calogero-Moser model. This corresponds to compactifications of the $D$-type $(2,0)$ theories \cite{Lemos:2012ph}. The $D_2$ model is just two copies of the Lam\'{e} system.}

\ 

We leave the analysis of expansion of indices in terms of eigenfunctions of this model for future work. We mention here that similarly to the $\alpha=1$ Schur index for ${\cal N}=2$ theories, the case of $g_i=1$ and $\lambda=1$  corresponds again to a  free fermionic particle limit here.

\

    \section{Summary and Discussion}\label{sec:summaryanddiscussion}

    We have discussed an interplay between the non-relativistic limit of some integrable models and special limits of the 4d supersymmetric index. In particular we have elaborated on  the relation of the generalized Schur index of ${\cal N}=2$ class ${\cal S}$ SCFTs and the eigenfunctions of the elliptic Calogero-Moser model (the non-relativistic limit of the Ruijsenaars-Schneider model). We have also discussed a novel limit of the supersymmetric index of the ${\cal N}=1$ $4d$ SCFTs obtained as Riemann surface compactifications of the rank $Q$ E-string theory and related them to the elliptic Inozemtsev model (the non-relativistic limit of the van~Diejen model). 

    In both models we have discussed, the Calogero-Moser and the Inozemtsev models, one has a many-particle quantum mechanical system with a potential. The potential has couplings related to ratios of chemical potentials of the $4d$ index computation. The eigenfunctions depend on the values of the couplings (one coupling in the Calogero-Moser model and $4(+1)$ couplings in the Inozemtsev model). Interestingly, switching off the couplings completely still leads to non-trivial and interesting indices though the underlying Hamiltonians are free. In fact there are two limits with vanishing couplings but with different orthogonality measures: the two choices can be identified as free bosons or free fermions. In particular for the Calogero-Moser model in the free fermionic limit one obtains precisely the Schur index.  In the Inozemtsev model the free fermionic limit leads to very similar expressions. In particular, we have explicitly seen in some examples that in fact some of the indices of E-string compactifications coincide with indices of class $\mathcal{S}$ $(2,0)$ compactifications. 

    When indices of two different theories are equal one should inquire whether there is a physical explanation. 
    We have reviewed that theories related by RG flows sometimes have same (restricted) indices. It is known that $6d$ SCFTs are interrelated by tensor branch flows. 
    In particular the E-string can be related by such a flow to a $(2,0)$ SCFT. This begs the question whether equalities of indices we observe can be explained by such flows across dimensions: flows in six dimensions with subsequent punctured and flux Riemann surface compactifications. For example, some relations of this type were studied in \cite{Razamat:2019mdt}. In particular, as we have observed also in this paper, in such across dimension flows compactifications of one SCFT on a surface of certain genus, flux, and number/types of punctures can flow to the same theory as another SCFT on a surface of same genus but possibly other values of flux and other numbers/types of punctures. Thus the generalized indices corresponding to non-relativistic limits could be a useful tool to study such flows. We leave exploring this possibility for  future research.\footnote{For a discussion of punctures in class ${\cal S}$ compactifications in terms of supergravity fluxes in the string theoretic setup, see \cite{Bah:2019jts}.}

    The results of the paper raise many additional  questions. Let us discuss several such curiosities. One  particular  question is whether for other types of compactifications of six dimensional SCFTs, well defined non-relativistic limits of the index exist. The ${\cal N}=1$ index depends in general on two parameters, $q$ and $p$. In non-relativistic limits $p$ is sent to one while $q$ is kept finite and generic. In order to obtain a finite limit some other parameters of the index corresponding to global symmetry need to be tuned on.  In case of RS model/$(2,0)$ compactifications this is the parameter $t$. In case of the van~Diejen model/E-string compactifications these are the fugacities for the Cartan generators of the $\E_8\times \SU(2)$ global symmetry of the model. Many other six dimensional $(1,0)$ SCFTs have global symmetries, and thus lead to integrable models with additional parameters. For example $ADE$  conformal matter theories \cite{DelZotto:2014hpa,Heckman:2015bfa} and their compactifications \cite{Gaiotto:2015usa,Razamat:2016dpl,Bah:2017gph,Kim:2018bpg,Kim:2018lfo,Kim:2023qbx}. Some of these integrable models were studied \cite{Gaiotto:2015usa,Maruyoshi:2016caf,Ito:2016fpl,Nazzal:2021tiu,Nazzal:2023bzu} and it will be interesting to understand their possible non-relativistic limits.
    
    On the other hand, some of the six dimensional SCFTs do not have global symmetry \cite{Seiberg:1996qx,Bershadsky:1997sb} and lead to $4d$ theories with no continuous global symmetries \cite{Razamat:2018gro}. Thus one obtains integrable models which depend only on $q$ and $p$ \cite{Razamat:2018zel,Ruijsenaars:2020shk,Nazzal:2023wtw}. For such models it seems that a non-relativistic limit does not exist. The integrable models here are given by elliptic difference operators with non-trivial interaction terms. However, taking a non-relativistic limit (sending say $p\to 1$ which can be thought of as sending the "speed of light" to infinity) also corresponds to taking the interactions to a singular limit. Thus in this respect, these integrable models are {\it intrinsically strongly coupled} and do not possess a natural free limit. Historically, the relativistic models were obtained as generalization of non-relativistic ones ({\it e.g.} the Ruijsenaars model is a generalization of the Calogero-Moser model). It would be interesting to understand whether the strongly-coupled models obtained in this way \cite{Razamat:2018zel,Ruijsenaars:2020shk,Nazzal:2023wtw}
    are an indication of existence of a large landscape of such models or rather an esoteric curiosity. 

    The fact that Schur indices of ${\cal N}=2$ SCFTs can be in some situations realized as limiting cases of a more general index (either generalized Schur for ${\cal N}=2$ or the Inozemtsev index, {\it etc}) is interesting also from another perspective. Over the years, a lot of surprising relations were discovered
    between Schur indices of different theories which do not have first principle derivations at the moment. For example, 
    indices of AD theories can be written in terms of matrix models with a ``wrong'' statistics, {\it e.g.} \cite{Buican:2017rya}, or indices of certain theories look like indices of other theories with parameters rescaled \cite{Buican:2019kba,Kang:2021lic}.\footnote{Maybe the simplest example is the fact that the Schur index of $D\left[\SU(2N+1)\right]$ AD theory is the same as that of adjoint half-hypermultiplet of $\SU(2N+1)$ with $q\to q^2$ \cite{Xie:2016evu}.} It would be interesting to understand whether some of these or similar observations can be explained once we consider the more general setup of non-relativistic indices of different classes of theories. Another example of such a question is given by observation  that generalized Schur indices can be extended to include additional continuous parameters following a relation to $2d$ CFT computations \cite{Chandra:2025qpv}.

    Finally, let us emphasize that it would be very interesting to understand better which, and how, physical information of the $4d$ SCFTs is encoded in  
    the non-relativistic limits of the index. For example, for general values of the coupling the index will have a $q$-expansion with non-integer couplings and thus does not a~priori correspond to a counting of states/operators problem. On the other hand for some values of the coupling one does recover expansions with integers and these hint at various physical insights \cite{Deb:2025ypl,Deb:2025ddc,Chandra:2025qpv}. A related point is that non-relativistic indices of theories with known Lagrangians are given as integrals of modular functions. It will be thus interesting to understand modularity properties of these quantities and the physical meaning of this modularity, see {\it e.g.} \cite{Razamat:2012uv,Beem:2017ooy,Gadde:2020bov,Beem:2021zvt,Pan:2021mrw,Pan:2025vyu,Pan:2024bne}. Such modularity properties could possibly be also related to further dimensional reductions to three and lower dimensions \cite{Dedushenko:2023cvd,ArabiArdehali:2024ysy}.

    \ 

    \noindent We leave these and other questions for future investigations.

\ 

\noindent{\bf Acknowledgments}:~
We are grateful to Leonardo Rastelli and Gabi Zafrir for insightful discussions.   
 The research of SSR and RBZ is supported in part by  the Planning and Budgeting committee, by the Israel Science Foundation under grant no. 2159/22, and by BSF-NSF grant no. 2023769. HK is supported by the National Research Foundation of Korea (NRF) grant funded by the Korean government (MSIT) (2023R1A2C1006542). The work of AD is supported in part by NSF grant PHY-2513893 and by the Simons Foundation grant 681267 (Simons Investigator Award).

    \appendix

\section{Comments on the Schur index and RG flows}\label{app:SchurWithMasses}

The Schur index a~priori is well defined also if the $\U(1)_r$ symmetry is broken. However, there are some subtleties which we will mention here.
As a trace over the states of the system the Schur index is defined as,
\be\label{eq:SchurIndex}
{\cal I}_q=\Tr_{{\mathbb S}^3}(-1)^F q^{\Delta-R}\, e^{-\beta(\Delta+(j_1-j_2)-2R)}\,.
\ee    The index is independent of $\beta$ which serves as a regulator for the computation of the trace. Using the notations of \cite{Gadde:2011uv} this index counts states annihilated by two supercharges, ${\cal Q}_{1+}$ and $\widetilde {\cal Q}_{1\dot{-}}$, which satisfy,
\be
2\{{\cal Q}_{1+},\,\left({\cal Q}_{1+}\right)^\dagger\}=\Delta+2j_1-2R-r\,,\qquad 
2\{\widetilde {\cal Q}_{1\dot{-}},\,\left(\widetilde{\cal Q}_{1\dot{-}}\right)^\dagger\}=\Delta-2j_2-2R+r\,.
\ee The charge coupling to the regulator thus is given by an anti-commutator of a combination of two supercharges,
\be
{\cal Q}\equiv \frac{{\cal Q}_{1+}+\widetilde {\cal Q}_{1\dot{-}}}{\sqrt{2}}\,,\qquad
2\{{\cal Q},\,{\cal Q}^\dagger\}=\Delta +(j_1-j_2)-2R\,.
\ee Thus assuming the general Witten index logic applies, if the index is finite it is expected  to be invariant of RG flows preserving symmetries for which we turn on fugacities. However, this is not strictly the case here. In order to argue rigorously that the index is invariant under the RG flow one needs to compute the index with a supercharge which is preserved along the flow. In particular as the superconformal symmetry is broken during the flow, one cannot use the conformal supercharges. This is important as in radial quantization the conformal supercharges are Hermitian conjugates of the usual supersymmetries. For general ${\cal N}=1$ supersymmetric index this problem   was analyzed in detail in \cite{Festuccia:2011ws}. This analysis goes through carefully defining indices as ${\mathbb S}^3\times {\mathbb S}^1$ partition function and analyzing the relevant supergravity background fields and preserved supersymmetries.  This analysis when performed for the Schur index in fact shows, surprisingly, that the index retains some information about relevant deformations such as masses \cite{DFZ}. Thus it is an invariant of RG flows 
breaking $\U(1)_r$ in much weaker sense. 

Let us discuss a simple example  of a free hypermultiplet $(Q,\widetilde Q)$ Schur index of which is given by,
\be \label{eq:SchurHyper}
{\cal I}_q^{(hyper)}(z)=\frac1{\theta_q(q^{\frac12}z)}\,.
\ee Here $z$ is fugacity for $\U(1)$ global symmetry. 
This index is not trivial. However, turning on mass deformation $W=m\widetilde Q Q$ preserves the symmetries which are used to compute this Schur index. The index of the IR theory should be equal to $1$, if the index is strict RG invariant, as the theory is gapped. Thus, we have a tension between these statements. The resolution is as stated above: although the index should be invariant of the RG flow the careful analysis of the index as a partition function reveals that some UV data is retained~ \cite{DFZ}. In particular this leads to the interesting relations of the Schur index of non-conformal theories and counting of massive BPS states on the Coulomb branch \cite{Cordova:2015nma,Cordova:2016uwk,Gaiotto:2024ioj,Gang:2024loa}.

The Schur index can be obtained from the full index \eqref{eq:FullIndex} by setting $t=q$. Here the regulator explicitly uses $\U(1)_r$ symmetry and thus there is no puzzle regarding invariance under RG flows. Let us discuss the free hypermultiplet from this point of view.  The full index of the free hypermultiplet is given by,
\be
{\cal I}^{(hyper)}_{p,q,t}(z)=\prod_{i,j=0}^\infty\frac{1-\frac{p q}{t^{\frac12}}p^iq^jz^{-1}}{1-t^{\frac12}p^iq^jz}\frac{1-\frac{p q}{t^{\frac12}}p^iq^jz}{1-t^{\frac12}p^iq^jz^{-1}}=\Gamma_e(t^{\frac12}z^{\pm1})\,.
\ee Setting here $t=q$ one obtains immediately \eqref{eq:SchurHyper},
\be 
{\cal I}^{(hyper)}_{p,q,q}(z)=\prod_{i,j=0}^\infty\frac{1-q^{\frac12}p^{i+1}q^jz^{-1}}{1-q^{\frac12}p^iq^jz}\frac{1-q^{\frac12}p^{i+1}q^jz}{1-q^{\frac12}p^iq^jz^{-1}}=\frac1{\theta_q(q^{\frac12}z)}\,.
\ee Note infinite number of terms in the numerator and denominator cancel against each other. The existence of fugacity $p$ is the natural regulator. We can turn on this fugacity if we have $\U(1)_r$ symmetry. Switching $p$ off and not turning on another regulator this limit is ill defined.

\section{Contributions of ${\cal N}=2$ multiplets in the non-relativistic limit}\label{app:multipletssection}

The non-relativistic limit of the index of an interacting ${\cal N}=2$ theory, such as a conformal gauge theory, has pole singularities due to non-trivial Coulomb branches. Stripping off this singularity leads to an expansion of the index with non-integer coefficients depending on $\alpha\in{\mathbb R}$. Nevertheless, one can ask what happens to indices of individual superconformal multiplets in the non-relativistic limit. We will discuss this question here. 

Let us mention then how different irreps of the ${\cal N}=2$ superconformal algebra contribute in the non-relativistic limit to the index. 
We follow the notations of \cite{Gadde:2011uv} and remind that in the non-relativistic limit the following holds,
\be
\sigma\,\tau \to e^{-\epsilon}\,,\qquad \sigma\rho=(\sigma\tau)^\alpha=e^{-\alpha\epsilon}\qquad\to\qquad \tau-\rho= \sigma^{-1}e^{-\epsilon}(1-e^{-\epsilon(\alpha-1)})\,.
\ee We obtain that different multiplets contribute as,
\be
&&{\cal I}_{\bar {\cal C}_{R,r(j_1,j_2)}}\to 0\,, \qquad 
{\cal I}_{\hat {\cal C}_{R(j_1,j_2)}} \to \alpha\, \frac{(-1)^{2(j_1+j_2)}\rho^{4+2R+2j_2+2j_1}}{1-\rho^2}\,,\qquad 
{\cal I}_{{\cal D}_{0(0,j_2)}} \to \alpha\,\frac{\rho^{2+2j_2}}{1-\rho^2}\,,\;\;\;\;\;\;\;\;\;\\
&&{\cal I}_{\bar {\cal E}_{r(j_1,0)}} \to (1-\alpha)\,(-1)^{2j_1}\rho^{-2j_1}\frac{1-\rho^{2+4j_1}}{1-\rho^2}\,,\qquad {\cal I}_{\bar {\cal D}_{0(j_1,0)}} \to (-1)^{2j_1}\rho^{-2j_1}\frac{1-\alpha-\rho^{2+4j_1}}{1-\rho^2}\,.\nonumber
\ee Note that the only possible $\bar {\cal D}$ and  $\bar {\cal E}$ multiplets in a Lagrangian theory have $j_1=0$.
This is also usually conjectured to be true for strongly coupled theories with no manifestly ${\cal N}=2$ supersymmetric Lagrangian.
Thus, these two multiplets in fact contribute,
\be 
{\cal I}_{\bar {\cal E}_{r(0,0)}} \to 1-\alpha\,,\qquad {\cal I}_{\bar {\cal D}_{0(0,0)}} \to \frac{1-\alpha-\rho^{2}}{1-\rho^2}\,.
\ee Note that contribution of any single multiplet is finite in the limit. However,
since contributions of $\bar {\cal E}_{r(0,0)}$ and $\bar {\cal D}_{0(0,0)}$ are order zero in $\rho$ (not suppressed by powers of $\rho$, or equivalently $q$), these can lead to divergences if there are infinite numbers of such multiplets. Indeed in the case of a non-trivial Coulomb branch such divergences do occur. 
We also note that for $\alpha=1$ we recover contributions of multiplets in the Schur limit while for $\alpha=0$ the Coulomb/mass limit.

\section{Definitions of special functions}\label{app:definitions}

Let us briefly summarize our definitions of basic special functions. We have the q-Pochhammer symbol,
\be
(z;q)=\prod_{\ell=0}^\infty (1-z q^\ell)\,,
\ee and the theta-function,
\be
\theta_q(z)=(z;q)(q\,z^{-1};q)=\prod_{\ell=1}^\infty(1-z q^\ell)(1-q^{1+\ell}z^{-1})\,.
\ee 
We also use the closely related definition,
\be
\theta_1(z|\tau)=-iz^{\frac{1}{2}}q^{\frac{1}{8}}(q;q)(z\, q;q)(z^{-1};q)\,,\qquad q=e^{2\pi i\tau}\,.
\ee
The elliptic Gamma function is defined to be,
\be
\Gamma_e(z)=\prod_{i,j=0}^\infty \frac{1-q^{i+1}p^{j+1}z^{-1}}{1-p^i q^j z}\,.
\ee We use the shorthand notation,
\be
f(z^{\pm1})=f(z^{+1})\,f(z^{-1})\,,\qquad f(\pm z)=f(+z)\,f(-z)\,.
\ee
        
\section{More technical details on $A_2$ wavefunctions}\label{app:A2technicaldetails}

In this appendix we give explicit expressions for the normalized eigenfunctions and structure constants for $A_2$ case. We will perform the computation of the index up to order $q^2$ and therefore various expressions are quoted to the relevant order of $q$. We have discussed how to obtain eigenfunctions using the $5d$ instanton computation in section \ref{sec:A2-instanton}. As a check on the procedure, we have also derived eigenfunctions directly by solving for the spectrum of the elliptic Calogero-Moser model in expansion in $q$ taking the leading contribution to be given by a Jack polynomial. The results given here were obtained by this second method and they agree with the $5d$ instanton computation.   
We have normalized the eigenfunctions defined in \eqref{eq:A2normalizedeigenfunctions} to be orthonormal.\footnote{We need the normalization for arbitrary real $\alpha$. To find it here we use a fitting procedure to find the normalization. We first evaluate the norm for various values of $\alpha$ and then fit it with a suitable function. We use the inbuilt command \texttt{FindSequenceFunction} in  \texttt{Mathematica}. This is very convenient for evaluation integrals as a function of $\alpha$ and we also use it to evaluate $\mathcal{M}_{\lambda_1,\lambda_2}$ (see equation \eqref{eq:mlambda}).} We parametrize the wavefunctions as follows, 
\begin{equation}
	\psi_{\lambda_1,\lambda_2}=\frac{\ppsi_{\lambda_1,\lambda_2}}{\sqrt{\cC_{\lambda_1,\lambda_2}}}~,
\end{equation} where,
\begin{equation}
    \begin{split}
	\ppsi_{0,0}(\mathbf{a})&=1+q\frac{3 \alpha^2 (\chi_{\mathbf{8}}-2)}{2 \alpha+1}
    +q^2\left(\frac{3 \alpha ^2 (3 \alpha +1) \chi _{\mathbf{27}}}{8 \alpha +4}+\frac{3 \alpha ^2 \left(3 \alpha ^2-2 \alpha -1\right) \left(\chi _{\bar{\mathbf{3}}}+\chi _\mathbf{3}\right)}{4 (\alpha +1) (2 \alpha +1)}\right.\\& \left.-\frac{6 \alpha ^2 \left(3 \alpha ^4+\alpha ^3-3 \alpha -1\right) \chi _\mathbf{8}}{(\alpha +1) (2 \alpha +1)^3}+\frac{3 \alpha ^2 \left(52 \alpha ^4+36 \alpha ^3-3 \alpha ^2-64 \alpha -21\right)}{4 (\alpha +1) (2 \alpha +1)^3}\right)+{\cal O}(q^3)\,,\nonumber
    \end{split}
\end{equation}
\begin{equation}
\begin{split}
	&\ppsi_{1,0}(\mathbf{a})=\ppsi_{1,1}(\mathbf{a}^{-1})=\nonumber\\&\;\;\;\;\chi_{\mathbf{3}}+q\left(\chi_{\overline{\mathbf{6}}}\frac{\alpha  \left(3 \alpha ^2+6 \alpha -1\right)}{2 (\alpha +1)^2}-\frac{5 \alpha ^3+10 \alpha ^2+5 \alpha +4}{2 (\alpha +1)^2}\chi_{\mathbf{3}}+\frac{3 \alpha ^2+\alpha }{2 \alpha +2}\chi_{\mathbf{15}}\right)+{\cal O}(q^2)\,.\nonumber
    \end{split}
\end{equation}
\begin{equation}
	\begin{split}
		\ppsi_{2,0}(\mathbf{a})&=\ppsi_{2,2}(\mathbf{a}^{-1})=\chi_{\bf 8}+\frac{2(\alpha-1)}{2\alpha+1}+q\left(\frac{\left(-3\alpha^6-3\alpha^5-\alpha^4+2\alpha^3+25\alpha^2+17\alpha+3\right)}{\alpha (\alpha+1)^3 (2\alpha+3)}\chi_{\overline{\mathbf{3}}}\right.\\&\left. +\frac{2\alpha \left(3\alpha^3+7\alpha^2+\alpha-1\right)
		}{(\alpha+1)^2 (2\alpha+3)}\chi_{\overline{\mathbf{15}}} -\frac{\left(3\alpha^6+25\alpha^5+45\alpha^4+48\alpha^3+53\alpha^2+23\alpha+3\right) }{\alpha (\alpha+1)^3 (2\alpha+3)}\chi_{\mathbf{6}}\right.\\& \left.
		+\frac{\alpha (3\alpha+2) }{2\alpha+3}\chi_{\overline{\mathbf{24}}}\right)+{\cal O}(q^2)\,,
	\end{split}
\end{equation}
\begin{equation}
	\begin{split}
		\ppsi_{2,1}(\mathbf{a})&=\chi_{\mathbf{8}}+\frac{2 (\alpha-1)}{2 \alpha+1}+
		q\left(\frac{-20 \alpha^6-48 \alpha^5+49 \alpha^4-4 \alpha^3+84 \alpha^2+62 \alpha+12}{\alpha (\alpha+2) (2 \alpha+1)^2 (2 \alpha+3)}\right.\\
        &\left.\frac{\alpha \left(3 \alpha^2+10 \alpha+2\right) }{(\alpha+2) (2 \alpha+3)}\chi_{\overline{\mathbf{10}}}+\frac{\left(4 \alpha^6-10 \alpha^5-98 \alpha^4-121 \alpha^3-125 \alpha^2-49 \alpha-6\right)}{\alpha (\alpha+2) (2 \alpha+1)^2 (2 \alpha+3)}\chi_{\mathbf{8}}\right.\\
        &
		\left.\frac{\alpha (3 \alpha+2)}{2 \alpha+3}\chi_{\mathbf{27}}+\frac{\alpha \left(3 \alpha^2+10 \alpha+2\right) }{(\alpha+2) (2 \alpha+3)}\chi_{\mathbf{10}}\right)+{\cal O}(q^2)\,.\nonumber
	\end{split}
\end{equation}
The normalization coefficients  $\cC_{\lambda_{\lambda_1,\lambda_2}}$ are computed to be as follows,
\begin{equation}
	\begin{split}
		\cC_{0,0}&=\frac{2\ 3^{-2+3 \alpha} \left(\frac{4}{3}\right)_{-1+\alpha}
			\left(\frac{5}{3}\right)_{-1+\alpha}}{\left((2)_{-1+\alpha}\right){}^2}-\frac{160\
			27^{-1+\alpha} q \left(\frac{7}{3}\right)_{-2+\alpha}
			\left(\frac{8}{3}\right)_{-2+\alpha}}{(1+2 \alpha)^2 (1)_{-2+\alpha}
			(2)_{-2+\alpha}}\\&+\frac{5\ 3^{-5+3 \alpha} (\alpha-1) \alpha^2 \left(-14-37 \alpha-41 \alpha^2-148
			\alpha^3-116 \alpha^4+32 \alpha^5\right) q^2 \left(\frac{7}{3}\right)_{-2+\alpha}
			\left(\frac{8}{3}\right)_{-2+\alpha}}{(1+2 \alpha)^4 \left((4)_{-2+\alpha}\right){}^2}\,,\nonumber
	\end{split}
\end{equation}
\begin{equation}
	\cC_{1,0}=\cC_{1,1}=\frac{2\ 3^{-1+3 \alpha} \left(\frac{4}{3}\right)_{-1+\alpha}
		\left(\frac{5}{3}\right)_{-1+\alpha}}{(1+2 \alpha)
		\left((2)_{-1+\alpha}\right){}^2}-\frac{3^{1+3 \alpha} \left(4+12 \alpha+\alpha^2+10 \alpha^3+5
		\alpha^4\right) q \left(\frac{1}{3}\right)_\alpha \left(\frac{2}{3}\right)_\alpha}{(1+\alpha)
		(1+2 \alpha) \left((2)_\alpha\right){}^2}\,,\nonumber
\end{equation}
\begin{equation}
	\begin{split}
		\cC_{2,0}&=\cC_{2,2}=\frac{2\ 3^{-1+3 \alpha} \left(\frac{5}{3}\right)_{-1+\alpha}
			\left(\frac{7}{3}\right)_{-1+\alpha}}{(1+2 \alpha)
			\left((3)_{-1+\alpha}\right){}^2}\nonumber\\&-\frac{4\ 3^{-1+3 \alpha} \left(9+66 \alpha+133 \alpha^2+69
			\alpha^3+56 \alpha^4+55 \alpha^5+12 \alpha^6\right) q \left(\frac{5}{3}\right)_{-1+\alpha}
			\left(\frac{7}{3}\right)_{-1+\alpha}}{\alpha (1+\alpha)^2 (1+2 \alpha) (3+2 \alpha)^2
			\left((3)_{-1+\alpha}\right){}^2}\,,\nonumber
	\end{split}
\end{equation}
\begin{equation}
	\begin{split}
		\cC_{2,1}&=-\frac{27^{1+\alpha} (2+\alpha) \left(-\frac{1}{3}\right)_{1+\alpha}
			\left(\frac{1}{3}\right)_{1+\alpha}}{(1+2 \alpha)^2
			\left((1)_{1+\alpha}\right){}^2}-4\ 3^{-2+3 \alpha} \,q\,\times\nonumber\\&\frac{ \left(36\!+\!336 \alpha\!+\!1105
			\alpha^2\!+\!1659 \alpha^3\!+\!1419 \alpha^4\!+\!682 \alpha^5\!+\!510 \alpha^6\!+\!288 \alpha^7\!+\!40 \alpha^8\right) 
			\left(\frac{5}{3}\right)_{\alpha-1} \left(\frac{7}{3}\right)_{\alpha-1}}{\alpha (1+2
			\alpha)^4 (3+2 \alpha)^2 (3)_{-1+\alpha} (4)_{-1+\alpha}}\,,\;\;\;\;\;\;\;\;\;\;\nonumber
	\end{split}
\end{equation}
\begin{equation}
	\cC_{3,0}=\cC_{3,3}=\frac{10\ 3^{-1+3 \alpha} \left(\frac{7}{3}\right)_{-1+\alpha}
		\left(\frac{8}{3}\right)_{-1+\alpha}}{(1+2 \alpha) (3+2 \alpha) (3)_{-1+\alpha} (4)_{-1+\alpha}}\,,\nonumber
\end{equation}
\begin{equation}
	\cC_{3,1}=\cC_{3,2}=\frac{5\ 3^{-1+3 \alpha} (3+\alpha) \left(\frac{7}{3}\right)_{-1+\alpha}
		\left(\frac{8}{3}\right)_{-1+\alpha}}{2 \alpha^2 (1+2 \alpha) (3+2 \alpha)
		\left((3)_{-1+\alpha}\right){}^2}\,,\nonumber
\end{equation}
\begin{equation}
	\cC_{4,0}=\cC_{4,4}=\frac{3^{2+3 \alpha} (1+\alpha) \left(\frac{4}{3}\right)_\alpha
		\left(\frac{5}{3}\right)_\alpha}{(1+2 \alpha) (3+2 \alpha) (3)_\alpha (4)_\alpha}\,,\nonumber
\end{equation}
\begin{equation}
	\cC_{4,1}=\cC_{4,3}=\frac{3^{7+3 \alpha} (4+\alpha) \left(-\frac{2}{3}\right)_{2+\alpha}
		\left(-\frac{1}{3}\right)_{2+\alpha}}{2 (1+2 \alpha) (3+2 \alpha)^2
		\left((1)_{2+\alpha}\right){}^2}\,,\nonumber
\end{equation}
\begin{equation}
	\cC_{4,2}=\frac{3^{2+3 \alpha} (3+\alpha) (4+\alpha) \left(\frac{4}{3}\right)_\alpha
		\left(\frac{5}{3}\right)_\alpha}{2 \left(3+5 \alpha+2 \alpha^2\right)^2
		\left((3)_\alpha\right){}^2}\,.
\end{equation}
Once the eigenfunctions are known, we can use the expansion of the free three punctured sphere, to obtain $\tilde{C}_{\lambda_1,\lambda_2}$. We can solve for $\tilde{C}_{\lambda_1,\lambda_2}$ by comparing both sides of equation \eqref{eq:A2hyptqftexpand} order by order in $q$ expansion. We rewrite $\tilde{C}_{\lambda_1,\lambda_2}$ as $\tilde{C}_{\lambda_1,\lambda_2}=\cC_{\lambda_1,\lambda_2}\CCC_{\lambda_1,\lambda_2}$ and obtain the following series expansion for $\CCC_{\lambda_1,\lambda_2}$,
\begin{equation}
	\begin{split}
		\CCC_{0,0}&=1+\tfrac{9 \alpha ^2 (4 \alpha +1) q}{(2 \alpha +1)^2}+\tfrac{12 \alpha ^3 q^{3/2}}{2 \alpha ^2+3 \alpha +1}\nonumber+\tfrac{3 \alpha ^2 \left(896 \alpha ^6+2648 \alpha ^5+2488 \alpha ^4+1370 \alpha ^3+895 \alpha ^2+308 \alpha +35\right) q^2}{4 (\alpha +1)^2 (2 \alpha +1)^4}\!+\!{\cal O}\!\left(\!q^{5/2}\!\right)\,,\nonumber
	\end{split}
\end{equation}
\begin{equation}
	\begin{split}
		\CCC_{1,0}\!=\!\CCC_{1,1}\!&=\!\alpha  \sqrt{q}\!+\!\tfrac{2 \alpha ^2 q}{\alpha +1}\!+\!\tfrac{\alpha  \left(22 \alpha ^4+55 \alpha ^3+34 \alpha ^2+23 \alpha +10\right) q^{3/2}}{2 (\alpha +1)^3}\nonumber \!+\!\tfrac{\alpha ^2 \left(28 \alpha ^4+63 \alpha ^3+30 \alpha ^2+27 \alpha +12\right) q^2}{(\alpha +1)^4}\!+\!{\cal O}\!\left(\!q^{5/2}\!\right)\,,\nonumber
	\end{split}
\end{equation}
\begin{equation}
	\begin{split}
		\CCC_{2,0}=\CCC_{2,2}&=\tfrac{1}{2} \alpha  (\alpha +1) q+\tfrac{2 \alpha ^3 q^{3/2}}{\alpha +1}\nonumber\\&+\tfrac{\left(48 \alpha ^8+389 \alpha ^7+1195 \alpha ^6+1881 \alpha ^5+1913 \alpha ^4+1608 \alpha ^3+1012 \alpha ^2+318 \alpha +36\right) q^2}{2 (\alpha +1)^2 (\alpha +2) (2 \alpha +3)^2}+{\cal O}\left(q^{5/2}\right)\,,\nonumber
	\end{split}
\end{equation}
\begin{equation}
	\begin{split}
		\CCC_{2,1}&=\alpha ^2 q+\tfrac{2 \alpha ^2 (2 \alpha +1) q^{3/2}}{\alpha +2}\nonumber\\&+\tfrac{\alpha  \left(176 \alpha ^8+1404 \alpha ^7+3720 \alpha ^6+4907 \alpha ^5+4638 \alpha ^4+4020 \alpha ^3+2594 \alpha ^2+744 \alpha +72\right) q^2}{(2 \alpha +1)^2 \left(2 \alpha ^2+7 \alpha +6\right)^2}+{\cal O}\left(q^{5/2}\right)\,,\nonumber
	\end{split}
\end{equation}
\begin{equation}
	\CCC_{3,0}=\CCC_{3,3}=\tfrac{1}{6} \alpha  \left(\alpha ^2+3 \alpha +2\right) q^{3/2}+\alpha ^3 q^2+{\cal O}\left(q^{5/2}\right)\,,\nonumber
\end{equation}
\begin{equation}
	\CCC_{3,1}=\CCC_{3,2}=\tfrac{2 \alpha ^4 q^{3/2}}{\alpha +1}+\tfrac{4 \alpha ^4 \left(3 \alpha ^3+11 \alpha ^2+8 \alpha +2\right) q^2}{(\alpha +1)^2 (\alpha +2) (\alpha +3)}+{\cal O}\left(q^{5/2}\right)\,,\nonumber
\end{equation}
\begin{equation}
	\CCC_{4,0}=\CCC_{4,4}=\tfrac{1}{24} \alpha  \left(\alpha ^3+6 \alpha ^2+11 \alpha +6\right) q^2+{\cal O}\left(q^{5/2}\right)\,,\nonumber
\end{equation}
\begin{equation}
	\CCC_{4,1}=\CCC_{4,3}=\tfrac{1}{6} \alpha ^2 \left(\alpha ^2+3 \alpha +2\right) q^2+{\cal O}\left(q^{5/2}\right)\,,\nonumber
\end{equation}
\begin{equation}
	\CCC_{4,2}=\tfrac{1}{4} \alpha ^2 (\alpha +1)^2 q^2+{\cal O}\left(q^{5/2}\right)\,.
\end{equation}
To obtain the expansion for the three punctured sphere corresponding to rank-one $\mathfrak{e}_6$ theory, we need to evaluate $C^{(A_2)}_{\lambda_1,\lambda_2}$. We rewrite the equation \eqref{eq:strucconstA2tqftexpand} as,
\begin{equation}
	\ccA^{(A_2)}_{\lambda_1,\lambda_2}=\frac{\CC_{\lambda_1,\lambda_2}^2}{\mathcal{M}_{\lambda_1,\lambda_2}}~
\end{equation}
where $\mathcal{M}_{\lambda_1,\lambda_2}$,
\begin{equation}
	\label{eq:mlambda}
	\mathcal{M}_{\lambda_1,\lambda_2}=\oint\frac{dz}{4\pi i z}\left(\theta(z^{\pm2};q)\right)^\alpha\frac{\psi_{\lambda_1,\lambda_2}(1,z,z^{-1})}{(q^{\frac12}z^{\pm1};q)^{2\alpha}}\,.
\end{equation}
Explicit evaluation of $\mathcal{M}_{\lambda_1,\lambda_2}$ using the quoted eigenfunctions are as follows,\footnote{We use the notation of the Pochhammer symbol as $(x)_n=\frac{\Gamma(x+n)}{\Gamma(x)}$. 
The expressions are quoted assuming $\alpha\geq 1$. Expressions for $\alpha$ outside of this range can be obtained by analytical continuation.}
\begin{equation}
	\begin{split}
		\mathcal{M}_{0,0}&=\frac{3 \sqrt{\tfrac{\left(\frac{16}{27}\right)^\alpha
					\left(\left(\frac{3}{2}\right)_{-1+\alpha}\right){}^2}{\left(\frac{4}{3}\right
					)_{-1+\alpha} \left(\frac{5}{3}\right)_{-1+\alpha}}}}{2 \sqrt{2}}+\frac{27 q
			\sqrt{\tfrac{\left(\frac{16}{27}\right)^\alpha \left(1+4 \alpha+4 \alpha^2-9 \alpha^3\right)^2
					\left(\left(\frac{5}{2}\right)_{-2+\alpha}\right){}^2}{(1+2 \alpha)^4
					\left(\frac{7}{3}\right)_{-2+\alpha} \left(\frac{8}{3}\right)_{-2+\alpha}}}}{4
			\sqrt{10}}\\&+18 \sqrt{2} q^{3/2} \sqrt{\tfrac{2^{4 \alpha} 3^{-3 \alpha} \alpha^6
				\left(\left(\frac{3}{2}\right)_{-1+\alpha}\right){}^2}{\left(1+3 \alpha+2
				\alpha^2\right)^2 \left(\frac{4}{3}\right)_{-1+\alpha}
				\left(\frac{5}{3}\right)_{-1+\alpha}}}\\&+\frac{1}{4} q^2
		\sqrt{\tfrac{\left(\frac{16}{27}\right)^\alpha \left(4+40 \alpha+164 \alpha^2+235 \alpha^3+74
				\alpha^4+571 \alpha^5-260 \alpha^6-1476 \alpha^7-648 \alpha^8\right)^2
				\left(\left(\frac{1}{2}\right)_\alpha\right){}^2}{(1+\alpha)^4 (1+2 \alpha)^8
				\left(\frac{1}{3}\right)_\alpha \left(\frac{2}{3}\right)_\alpha}}\,,\nonumber
	\end{split}
\end{equation}

\begin{equation}
	\begin{split}
		\mathcal{M}_{1,0}&=\mathcal{M}_{1,1}=\frac{3 \sqrt{\tfrac{\left(\frac{16}{27}\right)^\alpha
					\left(\frac{3}{2}\right)_{-1+\alpha}
					\left(\frac{5}{2}\right)_{-1+\alpha}}{\left(\frac{4}{3}\right)_{-1+\alpha}
					\left(\frac{5}{3}\right)_{-1+\alpha}}}}{2 \sqrt{2}}+3 \sqrt{2} \sqrt{q}
		\sqrt{\tfrac{\left(\frac{16}{27}\right)^\alpha \alpha^2
				\left(\frac{3}{2}\right)_{-1+\alpha} \left(\frac{5}{2}\right)_{-1+\alpha}}{(1+\alpha)^2
				\left(\frac{4}{3}\right)_{-1+\alpha}
				\left(\frac{5}{3}\right)_{-1+\alpha}}}\nonumber\\&+\frac{1}{2} q^2 \sqrt{\tfrac{3^{-1-3 \alpha}
				16^\alpha \left(-4-8 \alpha+9 \alpha^2+26 \alpha^3+9 \alpha^4\right)^2 \left(\frac{1}{2}\right)_\alpha
				\left(\frac{3}{2}\right)_\alpha}{(1+\alpha)^6 \left(\frac{1}{3}\right)_\alpha
				\left(\frac{2}{3}\right)_\alpha}}\,,\nonumber
	\end{split}
\end{equation}

\begin{equation}
	\mathcal{M}_{2,0}=\mathcal{M}_{2,2}=\sqrt{\tfrac{3^{-3 \alpha-1} 16^\alpha (\alpha+1)^2 \left(\frac{1}{2}\right)_\alpha \left(\frac{3}{2}\right)_\alpha}{\left(\frac{2}{3}\right)_\alpha \left(\frac{4}{3}\right)_\alpha}}\,,\nonumber
\end{equation}

\begin{equation}
	\mathcal{M}_{2,1}=\sqrt{\tfrac{2^{4 \alpha-3} 3^{1-3 \alpha} \alpha^2 (\alpha+2) \left(\left(\frac{3}{2}\right)_{\alpha-1}\right){}^2}{\left(\frac{5}{3}\right)_{\alpha-1} \left(\frac{7}{3}\right)_{\alpha-1}}}\,.
\end{equation} 
Combining everything together as in \eqref{eq:e6index} we obtained the result reported in \eqref{eq:e6generalzedSchurIndex}.

    \bibliographystyle{jhep}

	\bibliography{refs}

\end{document}